\documentclass[aps,twocolumn,rmp,epsfig]{revtex4}

\usepackage{epsf}
\usepackage{amsfonts}
\usepackage{amssymb}
\usepackage{graphicx}
\usepackage{color}
\usepackage{amsmath}
\usepackage[bookmarksnumbered, bookmarks, breaklinks, linktocpage]{hyperref}

\newcommand{\expect}[1]{\ensuremath{\left\langle#1\right\rangle}}

\newcommand{\diff}{\ensuremath{\mathrm{d}}}

\newcommand{\Sys}{\ensuremath{\mathcal{S}}}
\newcommand{\Env}{\ensuremath{\mathcal{E}}}
\newcommand{\Frag}{\ensuremath{\mathcal{F}}}

\newcommand{\Hh}{\ensuremath{H}}
\newcommand{\Hs}{\ensuremath{\Hh_{\Sys}}}

\newcommand{\Ham}{\ensuremath{\mathbf{{H}}}}

\newcommand{\Hsys}{\ensuremath{\Ham_{\mathrm{sys}}}}

\newcommand{\Msys}{\ensuremath{m_{\Sys}}}
\newcommand{\Menv}[1]{\ensuremath{m_{#1}}}
\newcommand{\omegasys}{\ensuremath{\omega_{\Sys}}}
\newcommand{\omegaenv}[1]{\ensuremath{\omega_{#1}}}
\newcommand{\omegabare}{\ensuremath{\Omega_0}}

\newcommand{\CC}[1]{\ensuremath{C_{#1}}}
\newcommand{\xsys}{\ensuremath{x_{\Sys}}}
\newcommand{\xenv}[1]{\ensuremath{y_{#1}}}
\newcommand{\psys}{\ensuremath{p_{\Sys}}}
\newcommand{\penv}[1]{\ensuremath{q_{#1}}}

\def\FCW{0.98\columnwidth}

\renewcommand{\omegasys}{\ensuremath{\omega_{\scriptscriptstyle\Sys}}}
\renewcommand{\Msys}{\ensuremath{m_{\scriptscriptstyle\Sys}}}

\newcommand{\bra}[1]    {\langle #1|}
\newcommand{\ket}[1]    {| #1 \rangle}
\newcommand{\bk}[2]     {\langle #1 | #2 \rangle}
\newcommand{\kb}[2]     {| #1 \rangle \! \langle #2 |}
\newcommand{\cH}        {{\mathcal H}}
\newcommand{\cS}        {{\mathcal S}}
\newcommand{\cA}        {{\mathcal A}}
\newcommand{\cE}        {{\mathcal E}}
\newcommand{\cN}        {{\mathcal N}}

\newcommand\cF{{\mathcal F}}
\newcommand\hocom[1]{}

\newcommand{\ba}{\begin{eqnarray}}
\newcommand{\ea}{\end{eqnarray}}
\newcommand{\bmath}{\begin{mathletters}}
\newcommand{\emath}{\end{mathletters}}
\newcommand{\ban}{\begin{eqnarray*}}
\newcommand{\ean}{\end{eqnarray*}}

\begin{document}

\title{Relative States and the Environment: Einselection, Envariance, Quantum Darwinism, and the Existential Interpretation}

\author{Wojciech Hubert Zurek}

\address{Theory Division, MS B213, LANL
    Los Alamos, NM, 87545, U.S.A.}

\date{\today}

\begin{abstract}
Starting with basic axioms of quantum theory we revisit  ``Relative State Interpretation'' set out 50 years ago by Hugh Everett III (1957a,b). His approach explains ``collapse of the wavepacket'' by postulating 
that observer perceives the state of the ``rest of the Universe'' {\it relative} to his own state, 
or -- to be more precise -- relative to the state of his records. This allows quantum theory to be universally valid. However, while Everett explains perception of collapse, relative state approach raises 
three questions absent in Bohr's Copenhagen Interpretation which relied on independent existence 
of an {\it ab intio} classical domain. One is now forced one to seek sets of preferred, effectively classical but ultimately quantum states that can define branches of the universal state vector, and allow observers to keep reliable records. Without such {\it (i) preferred basis} relative states are ``too relative'', and the approach suffers from {\it basis ambiguity}. Moreover, universal validity of quantum theory 
raises the issue of the {\it (ii) origin of probabilities, and of the Born's rule} $p_k = |\psi_k|^2$ which is simply postulated in textbook discussions. Last not least, even preferred quantum states (defined e.g. by the einselection -- environment - induced superselection) -- are still quantum. Therefore they cannot be found out by initially ignorant observers through direct measurement without getting disrupted. 
Yet, states of macroscopic object exist objectively and can be found out by anyone. So, we need to identify the {\it (iii) quantum origin of objective existence}. Here we show how mathematical structure of quantum theory supplemented by the only uncontroversial measurement axiom (that demands immediate repeatability -- and, hence, predictability -- of idealized measurements) 
leads to preferred sets of states: Line of reasoning reminiscent of the  ``no cloning theorem'' 
yields {\it (i) pointer states} which correspond to potential outcomes. Their stability is needed to establish effectively classical domain within quantum Universe, and to define  events such as measurement outcomes. This leads one to enquire about their probabilities or -- more specifically -- about the relation between probabilities of measurement outcomes and the underlying quantum state. We show that symmetry of entangled states -- {\it (ii) entanglement - assisted invariance} or {\it envariance} -- implies Born's rule. It also accounts for the loss of physical significance of local phases between Schmidt states. (in essence, for decoherence). Thus, loss of coherence between pointer states is a consequence of symmetries of entanglement (e.g., with the environment). It can be established without usual tools of decoherence (reduced density matrices and trace operation) that rely on Born's rule for physical motivation. Finally, we point out that monitoring of the system by the environment (process responsible for decoherence) will typically leave behind multiple copies of its pointer states. Only states that can 
survive decoherence can produce information theoretic progeny in this manner. This {\it (iii) quantum Darwinism} allows observers to use {\it environment as a witness} -- to acquire information about pointer states indirectly, leaving system of interest untouched and its state unperturbed. In conjunction with Everett's relative state account of the apparent collapse these advances illuminate relation of quantum theory to the classical domain of our experience. They complete {\it existential interpretation} based on the operational definition of objective existence, and justify our confidence in quantum mechanics as ultimate theory that needs no modifications to account for the emergence of the classical. 

\end{abstract}
\maketitle

\section{Introduction}

Quantum mechanics is often regarded as an essentially probabilistic theory, in which random 
collapses of the wavepacket governed by the rule conjectured by Max Born (1926) play a fundamental 
role (Dirac, 1958). Yet, unitary evolution dictated by Schr\"odinger's equation is deterministic. This 
clash of quantum determinism and quantum randomness is at the heart of interpretational controversies
reflected in the axioms that provide a textbook summary of quantum foundations:

(i) {\it State of a quantum system is represented by a vector in its Hilbert space} $\cH_{\cS}$.

(ii) {\it Evolutions are unitary (i.e., generated by Schr\"odinger equation).}

These two axioms imply, respectively, quantum principle of superposition and unitarity of quantum evolutions. They provide essentially complete summary of the formal structure of the theory. They 
seem to contain no premonition of either collapse or probabilities. However, in order to relate quantum 
theory to experiments one needs to establish correspondence between abstract state vectors in
$\cH_{\cS}$ and physical reality. The task of establishing this correspondence starts with the next axiom:

(iii) {\it Immediate repetition of a measurement yields the same outcome.}

Axiom (iii) is regarded as idealized (it is hard to devise in practice such non-demolition measurements,
but in principle it can be done). Yet -- as a fundamental postulate -- it is also uncontroversial. The very 
concept of a ``state'' embodies predictability which requires axiom (iii). The role of the state is to allow 
for predictions, and the most basic prediction is confirmation that the state is what it is known to be. 
However, in contrast to classical physics (where an unknown preexisting state
can be found out by an initially ignorant observer) the very next quantum axiom limits predictive 
attributes of the state:

(iv) {\it Measurement outcome is one of the orthonormal states -- eigenstates of the measured observable.}

This {\it collapse postulate} is the first truly controversial axiom in the textbook list. It is inconsistent with
the first two postulates: Starting from a general state $\ket {\psi_{\cS}}$ in a Hilbert space of the system
(axiom (i)), an initial state $\ket {A_0}$ of the apparatus $\cA$, and assuming unitary evolution (axiom (ii)) one is led to a superposition of outcomes:
$$\ket {\psi_{\cS}} \ket {A_0} =  (\sum_k a_k \ket {s_k}) \ket {A_0}
\Rightarrow
\sum_k a_k \ket {s_k} \ket {A_k}
\ , \eqno(1.1)$$
which is in apparent contradiction with axiom (iv).

 The impossibility to account -- starting with (i) and (ii) 
-- for the collapse to a single outcome postulated by (iv) was appreciated since Bohr (1928) and 
von Neumann (1932). It was -- and often still is -- regarded as an indication of the ultimate 
insolubility of the ``measurement problem''. It is straightforward to extend such insolubility demonstrations to various more realistic situations by allowing e.g. the state of the apparatus to be 
initially mixed. As long as axioms (i) and (ii) hold, one is forced to admit that the state of the Universe
after the measurement contains a superposition of many alternative outcomes rather than just one of
them as the collapse postulate (and our immediate experience) would have it.

Given this clash between mathematical structure of the theory and collapse (that captures the subjective 
impressions of what happens in real world measurements) one can proceed in two directions: One can 
accept -- with Bohr -- primacy of our immediate experience and blame the inconsistency of (iv) with 
the core of quantum formalism -- (i) and (ii) -- on the nature of the apparatus: According to Copenhagen 
Interpretation apparatus, observer, and, generally, any macroscopic object is classical: It does not abide 
by the quantum principle of superposition (that follows from (i)), and its evolution need not be unitary. 
So, axiom (ii) does not apply, and the collapse can happen on the border between quantum and classical. Uneasy coexistence of the quantum and the classical was a challenge to the unification instinct of physicists. Yet, it has proved to be surprisingly durable.

The alternative to Bohr's Copenhagen Interpretation and a new approach to the measurement problem was proposed by Hugh Everett III, student of John Archibald Wheeler, half a century ago. The basic idea is to abandon the literal view of collapse and recognize that a measurement (including appearance 
of the collapse) is already described by Eq. (1.1). One just need to include observer in the wavefunction, 
and consistently interpret consequences of this step. The obvious
problem -- ``Why don't I, the observer, perceive such splitting?'' -- is then answered by asserting that while the right hand side of Eq. (1.1) contains all the possible outcomes, the observer who recorded outcome \#17 will (from then on) perceive the Universe that is consistent with that random event reflected in his records. 
In other words, when global state of the Universe is $\ket \Upsilon$, and my state is
$\ket {{\cal I}_{17}}$, for me the state of the rest of the Universe ``collapses" to 
$\bk{{\cal I}_{17}} \Upsilon$. 

This view of the collapse is supported by axiom (iii);  
upon  immediate re-measurement the same state will be found.
Everett's (1957a) assertion:  ``The discontinuous jump into an eigenstate 
is thus only a relative proposition, dependent on the mode of decomposition of the total wave function 
into the superposition, and relative to a particularly chosen apparatus-coordinate value..." is consistent with quantum formalism:  In the superposition of Eq. (1.1) record state $\ket {A_{17}}$ can indeed imply detection of the corresponding state of the system, $\ket {s_{17}}$.  

Two questions immediately arise. First one concerns preferred states of the apparatus 
(or of the observer, or, indeed, of any object that becomes entangled with another quantum system). 
By the principle of superposition (axiom (i)) the state of the system or of the apparatus after 
the measurement can be written in infinitely many ways, each corresponding to 
one of the unitarily equivalent basis sets in the Hilbert spaces of the pointer of the apparatus 
(or memory cell of the observer). So;
$$
\sum_k a_k \ket {s_k} \ket {A_k}=\sum_k a'_k \ket {s'_k} \ket {A'_k} = \sum_k a''_k \ket {s''_k} \ket {A''_k} =... \eqno(1.2)
$$
This {\it basis ambiguity} is not limited to pointers of measuring devices (or cats, which in 
Schr\"odinger (1935) play a role of the apparatus). One can show that also very large systems 
(such as satellites or planets (Zurek, 1998a)) can evolve into very non-classical superpositions. In reality, this does not seem to happen. So, there is something that (in spite of the
egalitarian superposition principle enshrined in axiom (i)) picks out certain preferred quantum states, 
and makes them effectively classical. Axiom (iv) anticipates this. Before there is collapse, a set of preferred states one of which is selected by the collapse must be somehow chosen. There is nothing in 
writings of Everett that would hint at a criterion for such preferred states, and
nothing to hint that he was aware of this question. The second question concerns 
probabilities: How likely it is that -- after I measure $\cS$ -- I will become $\ket {{\cal I}_{17}}$?
Everett was very aware of its significance.

{\it Preferred basis problem} was settled by environment - induced superselection ({\it einselection}), usually discussed along with decoherence: As emphasized by Dieter Zeh (1970), apparatus, observers, and other macroscopic objects are immersed in their environments. This leads to monitoring of the system by its environment, described by analogy with Eq. (1.1). When this monitoring is focused on a specific observable of the system, its eigenstates form a {\it pointer basis} (Zurek, 1981): They entangle least with environment (and, therefore, are least perturbed by it). This resolves basis ambiguity. 

Pointer basis and einselection were developed and are discussed elsewhere (Zurek, 1982; 1991; 1993; 2003a; Paz and Zurek, 2001; Joos et al., 2003, Schlosshauer, 2004; 2007). However, they 
come at a price that might have been unacceptable to Everett:  Decoherence and einselection
usually employ reduced density matrices. Their physical significance derives from averaging, 
and is thus based on probabilities -- on Born's rule:

(v) {\it Probability $p_k$ of finding an outcome $\ket {s_k}$ in a measurement of a quantum system that was
previously prepared in the state $\ket \psi$ is given by $|\bk {s_k} \psi |^2$}.

Born's rule (1926) completes standard textbook discussions of the foundations of quantum theory. 
In contrast to collapse postulate (iv), axiom (v) is not in obvious contradiction with 
(i) - (iii), so one can adopt the attitude that Born's rule can be used 
to complete axioms (i) - (iii) and thus to justify preferred basis and symptoms of collapse 
via decoherence and einselection. This is the usual practice of decoherence (Zeh, 1970;
Zurek, 1991; 1998b; Paz and Zurek, 2001; Joos et al., 2003).

Everett believed that axiom (v) was inconsistent with the spirit of his approach. 
So, one might guess, he would not have been satisfied with the usual discussion of decoherence and 
its consequences. Indeed, he attempted to derive Born's rule from other quantum postulates. 
We shall follow his lead, although not his strategy which -- as is now known -- was flawed 
(DeWitt, 1971; Kent, 1990). 

One more axiom should added to postulates (i) - (v):

(o) {\it The Universe consists of systems.}

Axiom (o) is often omitted form textbooks as obvious. But, as pointed out by DeWitt (1970; 1971), it is 
useful to make it explicit in relative state setting where -- in contrast to Copenhagen Interpretation -- 
all of the Universe is quantum. As was noted before (Zurek, 1993; 2003a;
Schlosshauer, 2007), in absence of systems measurement problem disappears: Schr\"odinger 
equation provides a {\it deterministic} description of evolution of such an indivisible 
Universe, and questions about outcomes cannot be even posed. The measurement 
problem arises only because in quantum theory state of a collection of systems can evolve 
from a Cartesian product (where overall purity implies definite state of each subsystem) into an 
entangled state represented with a tensor product: The whole is definite and pure, but states 
of the parts are indefinite (and so there are no definite outcomes). 

This transition illustrated by Eq. (1.1) does not accord with our perception of what happens. We see 
a definite state of the apparatus in a Cartesian product with the corresponding state of the system. 
Relative state interpretation -- with observer included in the wavefunction -- restores the correspondence between equations and perceptions by what seems like a slight of hand: Everett decomposes  the global entangled tensor state into a superposition of branches -- Cartesian products -- labeled by 
observer's records. But we need more: Our goal is to understand the emergence of stable classical 
states from the quantum substrate, and origin of the rules governing randomness 
at the quantum-classical border. To this end in the next two sections we shall derive postulates (iv) 
and (v) from the non-controversial axioms (o)-(iii).  We shall then, in Section IV, account for
the ``objective existence'' of pointer states. This succession of results provides a quantum account of the classical reality. 

We start with the derivation of the preferred set of  {\it pointer states} -- the ``business end'' of the collapse postulate (iv) -- from axioms (o) - (iii). We will show that any set of states will do providing they are 
orthogonal. We will also see how these states are (ein)selected by the dynamics of the process of
information acquisition, thus following the spirit of Bohr's approach which emphasized the ability to
communicate results of measurements. Orthogonality of outcomes implies that the measured quantum observable must be Hermitean. We shall then compare this approach (obtained without resorting to reduced density matrices or any other appeals to Born's rule) with decoherence - based approach to pointer states and the usual view of einselection. 

Pointer states are determined by the dynamics of information transfer. They define outcomes  independently of the instantaneous reduced density matrix of the system (and, hence, of its initial state). Fixed outcomes define in turn events, and are key in discussion of probabilities in Section III. There 
we also take a fresh look at decoherence: It arises -- along with Born's rule -- from symmetries of 
entangled quantum states.

Given Born's rule and preferred pointer states one is still facing a problem: Quantum
states are fragile. Initially ignorant observer cannot find out an unknown quantum state
without endangering its existence: Collapse postulate means that selection of what 
to measure implies a set of outcomes. So, only a lucky choice of an observable could let 
observer find out a state without re-preparing it. The criterion for pointer states implied by axioms 
(o) - (iii) turns out to be equivalent to their stability under decoherence, and still leaves 
one with the same difficulty: How to find out effectively classical but ultimately quantum pointer state without re-preparing it? How to account for objective existence of ``classical reality'' using only  ``unreal'' quantum ingredients?  
 
The answer turns out to be surprisingly simple: Continuous monitoring 
of $\cS$ by its environment results in redundant records of in $\cE$. Thus, many observers 
can find out state of the system indirectly, from small fragments of the same $\cE$ that caused 
decoherence. Recent and still ongoing studies discussed in Section IV show how this replication selects  ``fittest'' states that can survive monitoring, and yield copious information-theoretic offspring: Quantum Darwinism favors pointer observables at the expense of their complements. Objectivity of preferred is quantified by their redundancy -- by the number of copies of the state of the system deposited in $\cE$.
Stability in spite of the environment is clearly a prerequisite for large redundancy.
Hence, pointer states do best in this information - theoretic ``survival of the fittest''.

Several interdependent steps account for the collapse, preferred states, probabilities, 
and objectivity -- for the appearance of  `the classical'.
It is important to take them in the right order, so that each step is based only on what is already established. This is our aim. Nevertheless, each section of this paper can be read separately:
Other sections are important to set the context, but are generally not required as a background.

\section{Einselection: Breaking of quantum unitary symmetry by cloning}

Unitary equivalence of all states in the Hilbert space -- the essence of axiom (i) -- is the basic symmetry of quantum theory. It is reaffirmed by axiom (ii) which admits only unitary evolutions. But unitarity is 
also a centerpiece of various ``proofs'' of insolubility of the measurement problem. Unitary 
evolution of a general initial state of a system $\cS$ interacting with the apparatus $\cA$ leads 
-- as seen in Eq. (1.1) -- to an entagled state of $\cS \cA$. Thus, in the end there is no single outcome
-- no collapse -- and an apparent contradiction with our experience. So -- the story 
goes -- measurement problem cannot be solved unless unitarity is somehow circumvented
(e.g., along the {\it ad hoc} lines suggested by the Copenhagen Interpretation).

We shall start with the same initial assumptions and follow similar steps, but arrive at a very different conclusion. This is because instead of the demand of a single outcome we shall only require that 
the results of the measurement can be confirmed (by a re-measurement), or communicated 
(by making a copy of the record). In either case one ends up with multiple copies of some state (of the 
system or of the apparatus). This ``amplification'' (implicit in axiom (iii)) calls for nonlinearity that would appear to be in conflict with the unitarity (and, hence, linearity) demanded by axiom (ii). 
Resolution of the tension between linear and non-linear demands brings to mind spontaneous symmetry breaking. As we shall see, amplification is possible, but only for a single orthogonal subset in 
the Hilbert space of the system. This conclusion (Zurek, 2007) extends the reach of {\it no cloning theorem}. It explains the need for quantum jumps.

This section shows -- on the basis of axioms (o) - (iii) and reasoning with a strong ``no cloning" flavor -- 
that quantum jumps (and hence at least Everettian collapse) are inevitable.
We shall also see how preferred Hermitian observable defined by the resulting orthogonal basis 
is related to the familiar  ``pointer basis''.

\subsection{Quantum origin of quantum jumps}

Consider a quantum system $\cS$ interacting with another quantum system $\cE$ (which can be 
for instance an apparatus, or as the present notation suggests, an environment). Let us suppose 
(in accord with axiom (iii)) that there is a set of states which  remain unperturbed by this interaction
-- e.g., that this interaction implements a measurement - like information transfer from $\cS$ to $\cE$:
$$ \ket {s_k} \ket {\varepsilon_0} \ \Longrightarrow \   \ket {s_k} \ket {\varepsilon_k} \ . \eqno(2.1) $$
We now show that a set of unperturbed states $\{  \ket {s_k} \}$ must be orthogonal providing that 
the evolution described above can start from an arbitrary initial  state $\ket {\psi_{\cS}}$ in $\cH_\cS$ 
(axiom (i)) and that it is unitary (axiom (ii)). From linearity alone we get:
$$\ket {\psi_{\cS}} \ket {\varepsilon_0} =  (\sum_k a_k \ket {s_k}) \ket {\varepsilon_0}
\Rightarrow
\sum_k a_k \ket {s_k} \ket {\varepsilon_k} = \ket {\Psi_{\cS\cE}}
\  . \eqno(2.2)$$

The total norm of the state must be preserved. 
After elementary algebra this leads to:
$$ 
Re \sum_{j,k} \alpha_j^* \alpha_k \bk {s_j} {s_k} = 
Re \sum_{j,k} \alpha_j^* \alpha_k \bk {s_j} {s_k} \bk {\varepsilon_j}{\varepsilon_k}
\  . \eqno(2.3)$$
This equality must hold for all states in $\cH_\cS$, and, in particular, for any phases of the complex 
coefficients $\{ \alpha_k \}$. Therefore, for any two states in the set $\{ \ket {s_k}\}$:
$$ \bk {s_j} {s_k} (1- \bk {\varepsilon_j}{\varepsilon_k}) \ = \ 0 \ . \eqno(2.4)$$
This equation immediately implies that $\{ \ket {s_k}\}$ must be orthogonal if they are to leave any 
imprint -- deposit any information -- in $\cE$ while remaining unperturbed: It can be satisfied
only when $\bk {s_j} {s_k} = \delta_{jk}$, unless $\bk {\varepsilon_j}{\varepsilon_k}=1$ -- unless 
states of $\cE$ bear no imprint of the states of $\cS$. 

In the context of quantum 
measurements Eq. (2.4) establishes the essence of axiom (iv) -- the orthogonality of outcome states. 
On other hand, when outcome states are orthogonal, any value of
$\bk {\varepsilon_j}{\varepsilon_k}$ is admitted, including $\bk {\varepsilon_j}{\varepsilon_k}=0$ 
which corresponds to a perfect record.

The necessity to choose between distinguishable (orthogonal) outcome states is then a direct 
consequence of the uncontroversial axioms (o) - (iii). It can be seen as a resolution of the tension
between linearity of quantum theory (axioms (i) and (ii)) and nonlinearity of the process of 
proliferation of information -- of amplification. This nonlinearity is especially obvious in cloning that 
in effect demands ``two of the same''. Our derivation above parallels proofs  of no-cloning 
theorem (Wootters and Zurek, 1982; Dieks, 1982; Yuen, 1986): The only difference -- in cloning 
copies must be perfect. Hence, scalar products must be 
the same, $\varsigma_{j,k}=\bk {\varepsilon_j}{\varepsilon_k}=\bk {s_j} {s_k}$. Consequently, 
we have a special case of Eq. (2.4):
 $$\varsigma_{j,k}(1-\varsigma_{j,k})=0 \ . \eqno(2.5)$$
 Clearly, there are only two possible solutions; $\varsigma_{j,k}=0$ (which implies orthogonality), or the trivial $\varsigma_{j,k}=1$. 
 
Indeed, we can deduce orthogonality of states that remain unperturbed while leaving small but distinct imprints in $\cE$ directly from the no-cloning theorem: As the states of $\cS$
remain unperturbed by assumption, arbitrarily many imperfect copies can be made. But each extra 
imperfect copy brings the collective state of all copies correlated with, say, $\ket {s_j}$, closer to orthogonality with the collective state of all of the copies correlated with any other state $\ket {s_k}$. 
Therefore, one could distinguish $\ket {s_j}$ for $\ket {s_k}$ by a measurement 
on a collection of sufficiently many of their copies with arbitrary accuracy, and, consequently, produce 
their ``clones''. So even imperfect copying (any value of $\bk {\varepsilon_j}{\varepsilon_k}$ except 1) 
that preserves the ``original" is prohibited by no-cloning theorem.

Similar argument based on unitarity was put forward in discussion of security of quantum cryptographic 
protocols. There, however, focus was on the ability to detect an eavesdropper through the perturbation 
she would have to inflict on the transmitted states (Bennett, Brassard, and Mermin, 1992), which is rather different from the quest for preferred basis considered here. 

Reader might be concerned that above discussion is based on idealized assumptions, which include 
purity of the initial state of $\cE$. Regardless of whether $\cE$ designates an environment 
or an apparatus, this is unlikely to be a good assumption. But this assumption is also easily bypassed:
A mixed state of $\cE$ can be always represented as a pure state of an enlarged system. This is the
so-called purification strategy: Instead of a density matrix 
$\rho_\cE=\sum_k p_k \ket {\varepsilon_k} \bra {\varepsilon_k}$ one can deal with a pure
entangled state of $\cE$ and $\cE'$: $\ket {\varepsilon_k \varepsilon_k'}=\sum_k \sqrt p_k \ket{\varepsilon_k} \ket {\varepsilon_k'}$ defined in $\cH_\cE \otimes \cH_{\cE'}$. So, when the initial
state of $\cE$ is mixed, there is always a pure state in an enlarged Hilbert space, and all of the steps
of the reasoning that lead to Eqs. (2.3) - (2.5) can be repeated leading to:
$$ \bk {s_j} {s_k} (1- \bk {\varepsilon_j \varepsilon_j'}{\varepsilon_k \varepsilon_k'}) \ = \ 0 \ , \eqno(2.6)$$
forcing one to the same conclusions as Eq. (2.3). 

Purification uses connection between pure states and density matrices by treating 
$\rho_\cE=\sum_k p_k \ket {\varepsilon_k} \bra {\varepsilon_k}$ as a result of a trace over some pure state. This appears to contradict our stated goal of deriving axiom (iii) without appeal to measurement 
axioms (iv) and (v): The connection $\rho_\cE$ with pure state $\ket {\varepsilon_k \varepsilon_k'}=\sum_k \sqrt p_k \ket{\varepsilon_k} \ket {\varepsilon_k'}$ does involve tracing and, hence, Born's rule. 
But there is a way to weaken this assumption: It suffices for to assume only that {\it some} such pure 
states in the enlarged Hilbert space exists. This does not rely on Born's rule, but it does assert that 
ignorance that is reflected in a mixed local state (here, of $\cE$) can be regarded as a consequence 
of entanglement, so that some pure global state of $\cE \cE'$ exists. So Born's rule is {\it not} needed 
for this purpose. But the existence of {\it some} such rule is needed. 

For a reader who is still suspicious of the procedure employed above we have an alternative: 
Unitary evolution preserves scalar products of density operators defined by $Tr\rho \rho'$. 
So, $Tr \kb {s_j} {s_j} \rho_{\cE} \kb {s_k} {s_k}  \rho_{\cE} = 
Tr \kb {s_j} {s_j} \rho_{{\cE}|j} \kb {s_k} {s_k}  \rho_{{\cE}|k}$, where $\rho_{{\cE}|j}$
and $\rho_{{\cE}|k}$ are mixed states of $\cE$ affected by the two states of $\cS$. This yields:
$$ |\bk {s_j} {s_k} |^2 (Tr \rho_{\cE}^2 - Tr\rho_{{\cE}|j} \rho_{{\cE}|k}) = 0 \eqno(2.7)$$
which can be satisfied only in the same two cases as before: Either $\bk {s_j} {s_k} =0$, or 
$Tr \rho_{\cE}^2 =Tr\rho_{{\cE}|j} \rho_{{\cE}|k}$ which implies (by Schwartz inequality) that 
$\rho_{{\cE}|j}= \rho_{{\cE}|k}$ (i.e., there can be no record of non-orthogonal states of $\cS$). 
This conclusion can be reached even more directly: It is clear that $\rho_{{\cE}|j}$ and $\rho_{{\cE}|k}$ 
have the same eigenvalues $p_m$ as $\rho_{\cE}=\sum_m p_m \kb {\varepsilon_m} {\varepsilon_m}$ from which they have unitarily evolved. Consequently, they differ from each other only in their eigenstates that contain record of the state of $\cS$, e.g.: $ \rho_{{\cE}|k}=\sum_m p_m \kb {\varepsilon_{m|k}} {\varepsilon_{m|k}}$. Hence, 
$Tr\rho_{{\cE}|j} \rho_{{\cE}|k}=\sum_m {p_m}^2 |\bk  {\varepsilon_{m|j}} {\varepsilon_{m|k}}|^2$, 
which coincides with $Tr {\rho_{\cE}}^2$ iff 
$|\bk {\varepsilon_{m|j}} {\varepsilon_{m|k}} |^2 =1$
whenever $p_m \neq 0$. So, $\rho_{{\cE}|j}=\rho_{{\cE}|k}$, and unless $\bk {s_j} {s_k} =0$, they can leave no record in $\cE$. 

Economy of our assumptions stands in contrast with the uncompromising
nature of our conclusions: Perfect predictability -- the fact that the evolution leading to information 
transfer preserves initial state of the system -- was, along with the principle of superposition, linearity 
of quantum evolutions, and preservation of the norm key to our derivation of inevitability of quantum jumps.

\subsection{Predictability killed the (Schr\"odinger) cat}

There are several equivalent ways to state our conclusions. To restate the obvious, we have 
established that outcome states of non-perturbing measurements must be orthogonal. This is the 
interpretation - independent part of axiom (iv) -- all of it except for the literal collapse. This is of course
enough for the Everettian relative state account of quantum jumps. So, a cat suspended between life
and death by Schr\"odinger (1935) is forced to make a choice between these two options because
these are the predictable options -- they allow (axiom (iii)) for confirmation (hence the above title).

Another way of stating our conclusion is to note that a set of orthogonal states
defines a Hermitian observable when supplemented with real eigenvalues. This is then a derivation of
the nature of observables. It justifies the textbook focus on the Hermitean operators sometimes invoked in an alternative statement of axiom (iv). 

We note that ``strict repeatability'' (that is, assertion that states $\{\ket {s_k}\}$ cannot change 
at all in courseof a measurement) is not needed: They can evolve providing that their scalar products
remain unaffected. That is:
 $$ 
 \sum_{j,k} \alpha_j^* \alpha_k \bk {s_j} {s_k} = 
 \sum_{j,k} \alpha_j^* \alpha_k \bk {\tilde s_j} {\tilde s_k} \bk {\varepsilon_j}{\varepsilon_k}. \eqno(2.8)$$
leads to the same conclusions as Eq. (2.2) providing that $\bk {s_j} {s_k}= \bk {\tilde s_j} {\tilde s_k}$.  
So, when $\ket  {\tilde s_j}$ and $\ket {\tilde s_k}$ are related with their progenitors 
by a transformation that preserves scalar product (e.g., by evolution in a closed system)
the proof of orthogonality goes through unimpeded. Both unitary and antiunitary transformations 
are in this class. 

We can also consider situations when this is not the case -- $\bk {s_j} {s_k}\neq \bk {\tilde s_j} {\tilde s_k}$. Extreme example of this arises when the state of the measured system retains no memory of 
what it was beforehand (e.g. $ \ket {s_j} \Rightarrow \ket 0, \ \ket {s_k} \Rightarrow \ket 0$). Then the apparatus can (and, indeed, by unitarity, has to) ``inherit'' information previously contained in the system. The need for orthogonality of $\ket {s_j}$ and $ \ket {s_k}$ disappears. 
Of course such measurements do not fulfill axiom (iii) -- they are not repeatable. For example, in quantum optics photons are usually absorbed by detectors, and coherent states play the role of the outcomes.

It is also interesting to consider sequences of information transfers involving many systems:
\begin{eqnarray}
\ket v \ket {A_0} \ket {B_0} \dots \ket {E_0} \Longrightarrow
\ket {\tilde v} \ket { A_v} \ket {B_0} \dots \ket {E_0} \Longrightarrow \dots \nonumber \\
\dots \Longrightarrow \ket {\tilde v} \ket {\tilde A_v} \ket {\tilde B_v} \dots \ket {E_v} \ . \nonumber \ \ \ \ \ \ (2.9a)
\end{eqnarray}
\begin{eqnarray}
\ket w \ket {A_0} \ket {B_0} \dots \ket {E_0} \Longrightarrow 
\ket {\tilde w} \ket { A_w} \ket {B_0} \dots \ket {E_0} \Longrightarrow \dots \nonumber \\
\dots \Longrightarrow \ket {\tilde w} \ket {\tilde A_w} \ket {\tilde B_w} \dots \ket {E_w} \ . \nonumber \ \ \  \ \ (2.9b)
\end{eqnarray}
Above we focused on a pair of states and simplified notation (e.g., $\ket v = \ket {s_k}, \ \ket w = \ket {s_k}$). Such ``von Nemann chains''  appear in discussion of quantum measurements (von Neumann, 1933), environment-induced decoherence (Zurek, 1982; 2003a), and -- as we shall see in Section IV -- in quantum Darwinism. As information about system is passed along this chain, links can be perturbed
(as indicated by {\it ``tilde''}). Unitarity implies that -- at each stage -- products of overlaps must 
be the same. Thus;
$$ \bra v \ket w = \bra {\tilde v} \ket {\tilde w} \bra {\tilde A_v} \ket {\tilde A_w} \bra {\tilde B_v} \ket {\tilde B_v} \dots \bra {E_v} \ket {E_w} \ , \eqno(2.10)$$
or -- after a logarithm of squares of the absolute values of both side is taken;
\begin{eqnarray}
 \ln |\bra v \ket w|^2 & = & \ln |\bra {\tilde v} \ket {\tilde w}|^2 + \ln| \bra {\tilde A_v} \ket {\tilde A_w}|^2 + \nonumber \\
& + & \ln | \bra {\tilde B_v} \ket {\tilde B_w}|^2 + \dots + \ln | \bra {E_v} \ket {E_w}|^2 .  \nonumber (2.11)
\end{eqnarray}
Therefore, when $\bra v \ket w \neq 0$, as the information about the outcome is distributed along 
the two von Neumann chains, quality of the records must suffer: Sum of logarithms above must equal 
$\ln |\bra v \ket w|^2$, and the overlap of two states is a measure of their distinguishability.
For orthogonal states there is no need for such deterioration of the quality of information; $\ln |\bra v \ket w|^2 =- \infty$, so arbitrarily many orthogonal records can be made. 

We emphasize that Born's rule was not used in the above discussion. The only two 
values of the scalar product that played key role in the proofs are ``0'' and ``1''. Both of them 
correspond to certainty; e.g. when we have asserted immediately below Eq. (2.4) that 
$\bk {\varepsilon_j}{\varepsilon_k}=1$ implies that these two states of $\cE$ are certainly identical. 
We shall extend this strategy of relying on certainty in the derivation of probabilities in Section III.

\subsection{Pointer basis, information transfer, and decoherence}

We are now equipped with a set of ``measurement outcomes'' or -- to put it in a way that ties in
with the study of probabilities we shall embark on in the Section III -- with a set of possible {\it events}. Our derivation above did not appeal to decoherence. However, symmetry breaking induced by decoherence yields einselection (which is, after all, due to information transfer to the environment). 
We will conclude that the two symmetry breakings are in effect two views of the same phenomenon.

Popular accounts of decoherence and its role in the emergence of the classical often start from the
observation that when a quantum system $\cS$ interacts with some environment $\cE$ 
``phase relations in $\cS$ are lost''. This is a caricature, at best incomplete if not misleading: 
It begs the question: ``Phases between {\it what}?''. This in turn leads directly to the main issue addressed by einselection: ``{\it What is the preferred basis}?". This question is often muddled in 
``folklore'' accounts of decoherence.

The crux of the matter -- the reason why interaction with the environment can impose classicality -- 
is precisely the emergence of the preferred states. Its role and the basic criterion for singling out 
preferred pointer states was discovered when the analogy between the role of the environment in
decoherence and the role of the apparatus in measurement were understood: What matters is that 
there are interactions that transfer information and yet leave selected states of the system unaffected. This leads one to einselection -- to the environment induced superselection of preferred pointer states
(Zurek, 1981). 

Our discussion showed that simple idea of preserving a state while transferring the information 
about it -- also the central idea of einselection -- is very powerful indeed. It leads to breaking 
of the unitary symmetry and singles out preferred pointer states without any need to involve usual tools 
of decoherence. This is significant, as reduced density matrices and partial trace employed in decoherence calculations invoke Born's rule, axiom (v), which relates states vectors and probabilities. 

To consider probabilities it is essential to identify outcomes separately from these probabilities 
(and, therefore, independently from the amplitudes present in the initial state of the measured system). 
Both einselection implemented via the predictability sieve (Zurek, 1993; Zurek, Habib, and Paz, 1993;
Paz and Zurek, 2001) and the ``no cloning'' approach above accomplish this goal. 

To compare derivation of preferred states in decoherence with their emergence from 
symmetry breaking imposed by axioms (o) - (iii) we return to Eq. (2.2). We also temporarily 
suspend prohibition on the use of partial trace to compute reduced density matrix of the system:
$$ \rho_\cS = \sum_{j,k} \alpha_j\alpha_k^* {\bk {\varepsilon_j} {\varepsilon_k}} \ket {s_j} \bra {s_k}
= Tr_{\cE} \kb {\Psi_{\cE\cS}}  {\Psi_{\cE\cS}} \ . \eqno(2.12)$$
Above we wrote $\rho_\cS$ in the pointer basis defined by its resilience in spite of the monitoring
by $\cE$. Resilience is the essence of the original definition of pointer states and einselection 
(Zurek, 1981; 1982). We note that pointer states will be in general different from Schmidt states
of $\cS$ -- eigenstates of $\rho_\cS$. They will coincide with Schmidt states of $\ket {\Psi_{\cE\cS}}=\sum_k a_k \ket {s_k} \ket {\varepsilon_k}$ only when $\{\ket{\varepsilon_k} \}$ -- their records in $\cE$ -- are orthogonal. We did not need such perfect orthogonality of $\{\ket{\varepsilon_k} \}$ to prove orthogonality of pointer states earlier in this section.

\hocom{So, a piece of decoherence ``folklore'' -- statement that ``decoherence causes reduced density 
matrix to be diagonal'' is shown to be incorrect. The error is mathematical and 
obvious: $\rho_\cS$ is Hermitian, so it is always diagonalized by the Schmidt states of $\cS$. }

``Folklore'' often assigns classicality to eigenstates of $\rho_{\cS}$. This identification is occasionally 
supported by some of decoherence proponents (Albrecht, 1992; Zeh, 1990; 2007) and taken for granted 
by others (e.g. Deutsch, 1986), but by and large it is no longer regarded as viable 
(see e.g. Schlosshauer, 2004; 2007): Eigenstates of $\rho_\cS$ are {\it not} stable. They depend 
on time and on the initial state of $\cS$, which disqualifies them as events in the sense of probability theory, as the ``elements of classical reality". A related problem arises in a very long time limit, when equilibrium sets in, so energy eigenstates diagonalize $\rho_{\cS}$. 

\hocom{At first, dependence on initial states 
does not sound like a problem -- after all, the outcome of a measurement should depend on the initial 
state of the system. But this is not the sort of dependence that corresponds to what happens in the
laboratory: Here the {\it set of possible outcomes} depends of the initial state of $\cS$, illustrating 
basis ambiguity we have pointed out earlier.}

As is frequently the case with folk wisdom, a grain of truth is nevertheless reflected in such 
oversimplified ``proverbs'': When environment acquires perfect knowledge of the states it 
monitors without perturbing and ${\bk {\varepsilon_j} {\varepsilon_k}}=\delta_{jk}$, pointer states 
``become Schmidt'', and end up on the diagonal of $\rho_\cS$. Effective decoherence assures 
such alignment of Schmidt states with the pointer states. Given that decoherence is -- at least 
in the macroscopic domain -- very efficient, this can happen essentially instantaneously. 
Still, this coincidence should not be used to attempt a redefinition of pointer states 
as instantaneous eigenstates of $\rho_{\cS}$ -- instantaneous Schmidt states. As we have 
already seen, and as will become even clearer in the rest of this paper, it is important to 
distinguish process that fixes preferred pointer states (that is, dynamics of information transfer 
that results in measurement as well as decoherence, but does not depend on the initial state 
of the system) from the probabilities of these outcomes (that are determined by this initial state). 

\subsection{Summary: The origin of outcome states}

Preferred states of quantum systems emerge from dynamics. Interaction between the system and 
the environment plays crucial role: States that are immune to monitoring by the 
environment are predictable, and at least in that sense the most classical.

Selection of pointer states is determined by the evolution -- i.e., in practice by the completely 
positive map (CPM) that represents open system dynamics. Therefore, instantaneous eigenstates 
of the reduced density matrix of the open systems will in general not coincide with the preferred pointer 
states singled out by einselection. However, as pointer states are capable of evolving predictably 
under the CPM in question, they will be -- eventually, and after decoherence has done its job, but before 
the system has equilibrated -- found on or near diagonal of $\rho_{\cS}$. It is nevertheless important to emphasize that definition of pointer states, e.g. -- via predictability sieve -- is based on stability, and {\it not} on this coincidence.

The lesson that derives from this section -- as well as from earlier studies, including the original 
definition of pointer states (Zurek, 1981) -- is that the preferred, effectively classical basis has nothing 
to do with the initial state of the system. This is brings to mind Bohr's insistence that the measured 
observable of the system is defined by the classical apparatus, so that it arises outside of the quantum 
domain. Here preferred observables are defined not by the ``classical apparatus'', but by the open 
system dynamics. Completely positive map describing it will preserve (at least approximately) certain 
subset of all the possible states. Extremal points of this set are projection operators corresponding 
to pointer states. In Bohr's view as well as here they are defined by something else than just the preexisting quantum 
states. We have defended the idea that they are classical
by pointing out to their predictability. In Section IV we shall show that they can be often found out by
observers without getting disrupted (a telltale sign of objective existence, which will reinforce case for their classicality).
 
In the next section we derive Born's rule. We build on einselection, 
but do it in a way that does not rely on axiom (v). In particular, use of reduced 
density matrices or completely positive maps we allowed in the latter part of this section
shall be prohibited. We shall use them again in section IV, only after Born's rule has been derived.

\section{Probabilities and Born's rule from the symmetries of entanglement}

The first widely accepted definition of probability was codified by Laplace (1820): When there are $N$ 
possible distinct outcomes and observer is ignorant of what will happen, all alternatives appear 
equally likely. Probability one should assign to any one outcome is then $1/N$. Laplace justified this 
{\it principle of equal likelihood} using {\it invariance} encapsulated in his `principle of indifference': 
Player ignorant of the face value of cards in front of him (Fig. 1a) will be {\it indifferent} when they are swapped before he gets the card on top, even when one 
and only one of the cards is favorable (e.g., a spade he needs to win).

\begin{figure*}[tb]
\begin{tabular}{l}
\vspace{-0.15in} 
\includegraphics[width=4.0in]{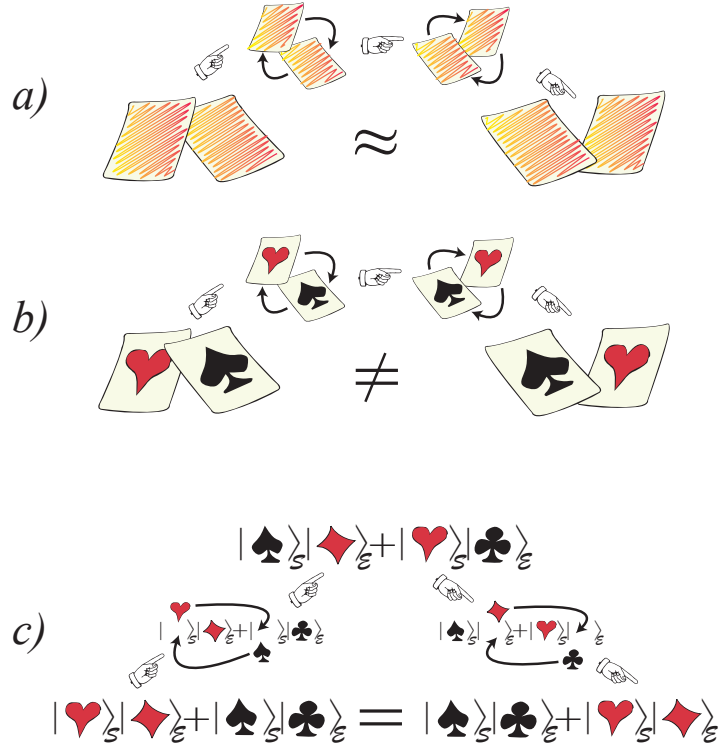}\\
\end{tabular}
\caption{{\it Probabilities and symmetry}: 
{\bf (a)} Laplace (1820) appealed to subjective invariance associated with `indifference' of the observer based on his ignorance of the real physical state to define probability through his {\it principle of equal likelihood}. When observers ignorance means he is indifferent to swapping (e.g., of cards), alternative 
events should be considered equiprobable. So, for the cards above, subjective probability 
$p_\spadesuit ={ \frac 1 2}$ would be inferred by an observer who does not know their face value, 
but knows that one (and only one) of the two cards is a spade.
{\bf (b)} The real physical state of the system is however altered by the swap -- it is not `indifferent',
illustrating subjective nature of Laplace's approach. The subjectivity of equal likelihood probabilities 
posed foundational problems in statistical physics, and led to the introduction of imaginary ensembles. 
They were fictitious and  subjective, but their state could be associated with probabilities 
defined through relative frequency -- an objective property (albeit of a fictitious infinite ensemble).
{\bf (c)} Quantum theory allows for an objective definition of probabilities based on {\it a perfectly known state} of a composite system and symmetries of entanglement: When two systems ($\cS$ and $\cE$) 
are maximally entangled (i.e, Schmidt coefficients differ only by phases, as in the Bell state
above), a swap $\ket \spadesuit \bra \heartsuit + \ket \heartsuit \bra \spadesuit$ in $\cS$ can be undone by `counterswap' $\ket \clubsuit \bra \diamondsuit + \ket \diamondsuit \bra \clubsuit$ in $\cE$. So, 
as can be established more carefully (see text),  $p_\spadesuit = p_\heartsuit=\frac 1 2$ follows from 
an objective symmetry of entanglement. This entanglement - assisted invariance ({\it envariance}) 
also causes decoherence of Schmidt states, allowing for additivity of probabilities of the effectively classical pointer states. Probabilities derived through envariance quantify indeterminacy of the state of 
$\cS$ alone given the global entangled state of $\cS\cE$. Complementarity between 
global observables with entangled eigenstates (such as $\ket {\bar \psi_{\cal SE}}$, Eq. (3.5)) and 
local observables (such as $\sigma_{\cS} = \sum_k \varsigma_k |s_k \rangle \langle s_k|\otimes {\bf 1}_{\cE}$) is reflected in the commutator $[\ket {\bar \psi_{\cal SE}}\bra {\bar \psi_{\cal SE}}, ~\sigma_{\cS} \otimes {\bf 1}_{\cE}]=\sum_k \varsigma_k (\ket {\bar \psi_{\cal SE}}\bra {s_k } \bra {\varepsilon_k} - h.c.$. 
It does not vanish, implying quantum indeterminacy responsible for the uncertainty about the outcome 
of a future measurement of $\sigma_{\cS} = \sum_k \varsigma_k |s_k\rangle \langle s_k|$ on $\cS$ alone.}
\label{cards}
\end{figure*}

Laplace's invariance under swaps captures {\it subjective} symmetry: Equal likelihood is a statement 
about observers `state of mind'  (or, at best, his records), and {\it not} a measurable property of the real 
physical state of the system which is altered by swaps (Fig. 1b). In classical setting probabilities defined 
in this manner are therefore ultimately unphysical. Moreove, indifference, likelihood, and probability are all ill-defined attempts to quantify the same ignorance. Expressing one undefined concept in terms of another undefined concept is not a definition.

It is therefore no surprise that equal likelihood is no longer regarded as a sufficient 
foundation for classical probability, and several other attempts are vying for primacy (Fine, 1973). 
Among them, relative frequency approach has perhaps the largest following, but attempts to use it in the relative states context (Everett,~1957a,b; DeWitt,~1970; 1971; Graham,~1973; Geroch, 1984) have been 
found lacking. This is because counting many world branches does not suffice. ``Maverick''  branches 
that have frequencies of events very different from these predicted by Born's rule are also a part of 
the universal state vector, and on the basis of relative frequencies alone there is no good reason to 
claim that observer is unlikely to be found on such a branch. To get rid of them one would have to assign 
to them -- without physical justification -- vanishing probabilities related to their small amplitudes. 
This goes beyond relative frequency approach, in effect requiring -- in addition to frequencies -- 
another measure of probability. Papers mentioned above introduce it {\it ad hoc}. This is consistent 
with Born's rule, but circular (Stein, 1984; Kent, 1990; Joos, 2000). Indeed, formal attempts based on the 
``frequency operator'' lead to mathematical inconsistencies (Squires, 1990). 
 
The problem can be ``made to disappear'' -- coefficients of maverick branches become small 
(along with coefficients of {\it all} the branches) -- in a limit of {\it infinite} and fictitious (and, hence, subjectively assigned) ensembles is introduced (Hartle,~1968; Farhi, Goldstone, and Guttman,~1989). 
Such infinite ensembles -- one might argue -- are always required by the frequentist approach also in 
the classical setting, but this is a weak excuse (see Kent, 1990). Moreover, in quantum mechanics 
they may pose problems that have to do with the structure of infinite Hilbert spaces (Poulin, 2005; Caves and Shack, 2005). It is debatable whether these mathematical problems are fatal, but it is also difficult to
disagree with Kent (1990) and Squires (1990) that the need to go to a limit of infinite ensembles 
to define probability in a finite Universe disqualifies them in the relative states setting. 

The other way of dealing with this issue is to modify physics so that branches with small enough 
amplitude simply do not count
(Buniy, Hsu, and Zee, 2006). We
believe it is appropriate to regard such attempts primarily as illustration of seriousness of the problem
at least until experimental evidence for the required modifications of quantum theory is found.

Kolmogorov's approach -- probability as a measure (see e.g. Gnedenko, 1968) --  bypasses the question 
we aim to address: How to relate probability to states. It only shows that any sensible assignment
(non-negative numbers that for a mutually exclusive and exhaustive set of events sum up to 1) will do. And it assumes additivity of probabilities while quantum theory provides only additivity of complex 
amplitudes (and, typically, these two additivity requirements are mutually inconsistent, as double slit experiment famously demonstrates). 

Gleason's theorem (Gleason, 1957) implements axiomatic approach to probability by looking for an additive measure on Hilbert spaces. It leads to Born's rule, but provides no physical insight into why the result should be regarded as probability. Clearly, it has not settled the issue: Rather, it is often cited (Hartle,~1968; Graham,~1973; Farhi, Goldstone, and Guttman,~1989; Kent,~1990) as a motivation 
to seek a physically justified derivation of Born's rule.

We shall now demonstrate how quantum entanglement leads to probabilities based on a symmetry, but 
-- in contrast to subjective equal likelihood -- on {\it objective} symmetry of {\it known} quantum states.

\subsection{Envariance}

A pure entangled state of 
a system $\cS$ and of another system (which we call ``an environment $\cE$'', anticipating 
connections with decoherence) can be always written as:
$$
|\psi_{\cal SE}\rangle = \sum_{k=1}^N a_k |s_k\rangle |\varepsilon_k\rangle
\ .  \eqno(3.1)
$$
Here $a_k$ are complex amplitudes while $\{\ket {s_k}\}$ and  $\{\ket {\varepsilon_k}\}$ are orthonormal bases in the Hilbert spaces $\cH_\cS$ and $\cH_\cE$. This {\it Schmidt decomposition} of pure 
entangled $|\psi_{\cal SE}\rangle$ is a consequence of a theorem of linear algebra that predates
quantum theory. 

Schmidt decomposition demonstrates that any pure entangled bipartite state is
a superposition of {\it perfectly correlated outcomes} of judiciously chosen measurements 
on each subsystem: Detecting $\ket {s_k}$ on $\cS$ implies, with certainity, $\ket {\varepsilon_k}$ 
for $\cE$, and {\it vice versa}. 

Even readers unfamiliar with Eq.~(3.1) have likely relied on its consequences: Schmidt basis 
$\{ |s_k\rangle \}$ appears on the diagonal of the reduced density matrix 
$\rho_{\cal S}= Tr_{\cE}|\psi_{\cal SE}\rangle \langle \psi_{\cal SE}|$. But tracing presumes Born's rule 
we are trying to derive (see e. g. Nielsen and Chuang, 2000, for how Born's rule is used to 
justify physical significance of reduced density matrices). Therefore, we shall not use $\rho_{\cal S}$ as this could introduce circularity. Instead, we shall derive Born's rule from symmetries of $|\psi_{\cal SE}\rangle$.

\hocom{Somewhat surprisingly, Everett did not seem to know of the Schmidt decomposition. However, he did derive it  ``from scratch" (and called it ``canonical'') in his dissertation (Everett, 1957a).  Yet, it did 
not seem important enough to him deserve mention in his paper (Everett, 1957b). }

Symmetries reflect invariance. Rotation of a circle by an arbitrary angle, or
of a square by multiples of $\pi/2$ are familiar examples. Entangled quantum states 
exhibit a new kind of symmetry -- {\it entanglement - assisted invariance}
or {\it envariance}: When a state $|\psi_{\cal SE}\rangle$ of a pair ${\cal S,~E}$
can be transformed by $U_{\cal S}=u_{\cal S} \otimes {\bf 1}_{\cal E}$ acting
solely on ${\cal S}$, 
$$ 
 U_{\cal S}|\psi_{\cal SE}\rangle  =
(u_{\cal S} \otimes {\bf 1}_{\cal E})|\psi_{\cal SE}\rangle  =
|\eta_{\cal SE}\rangle \  ,  \eqno(3.2)
$$
but the effect of $U_{\cal S}$ can be undone by acting
solely on ${\cal E}$ with an appropriately chosen $U_{\cal E}=
{\bf 1}_{\cal S} \otimes u_{\cal E}$: 
$$
U_{\cal E}|\eta_{\cal SE}\rangle  =
({\bf 1}_{\cal S} \otimes u_{\cal E}) |\eta_{\cal SE}\rangle
= |\psi_{\cal SE}\rangle \  ,  \eqno(3.3)
$$
then $|\psi_{\cal SE}\rangle$ is called envariant under $U_{\cal S}$ (Zurek, 2003a;b).

Envariance can be seen on any entangled $|\psi_{\cal SE}\rangle$. 
Any unitary operation diagonal in Schmidt basis $\{\ket {s_k}\}$:
$$u_{\cal S}=\sum_{k=1}^N \exp(i \phi_k)|s_k\rangle \langle s_k| \ , 
\eqno(3.4a)$$
is envariant: It can be undone by a {\it countertransformation}:
$$u_{\cal E}=\sum_{k=1}^N \exp(-i \phi_k)|\varepsilon_k\rangle
\langle \varepsilon_k|  \  , \eqno(3.4b)$$
acting solely on environment. 

In contrast to familiar symmetries (when a transformation has no effect on a state or an object) 
envariance is an {\it assisted symmetry}: The global state of $\cS \cE$ {\it is} transformed 
by $U_{\cal S}$, but it can be restored by acting on ${\cal E}$, physically distinct (e.g., spatially 
separated) from ${\cal S}$. When the state of $\cS \cE$ is envariant under 
some $U_{\cS}$, the state of $\cS$ alone must be obviously invariant under it.

Entangled state might seem an unusual starting point for the discussion of probabilities. After all, 
textbook formulation of Born's rule begins with a pure state. However, in 
Everett's approach entangled quantum state is a model of a measurement, with the outcomes corresponding to different states of the apparatus pointer (or of the memory of the observer). 
A similar sort of entanglement happens in course of decoherence. So it is natural to enquire about 
the symmetries of such states. And as we shall see below, 
envariance allows one to reassess the role of the environment and the origin of decoherence.

\subsection{Decoherence as a result of envariance}

Envariance of entangled states leads to our first conclusion: Phases of Schmidt coefficients are 
envariant under local (Schmidt) unitaries, Eqs. (3.4). Therefore, when a composite system is 
in an entangled state $|\psi_{\cal SE}\rangle$, the state of $\cS$ (or ${\cE}$) 
alone is {\it invariant under the change of phases} of $a_k$. In other words, the {\it state} of ${\cS}$
(understood as a set of all measurable properties of $\cS$ alone) cannot depend on
phases of Schmidt coefficients: It can depend only on their absolute values and on 
the outcome states -- on the set of pairs $\{ |a_k|, \ket{s_k} \}$. In particular (as we shall see
below) probabilities cannot depend on these phases.

So loss of phase coherence between Schmidt states -- decoherence -- is a consequence of envariance: Decoherence is, after all, selective loss of relevance of phases for the state of $\cS$. 
We stumbled here on its essence while exploring an unfamiliar territory, without the usual backdrop 
of dynamics and without employing trace and reduced density matrices. This encounter 
is a good omen: Born's rule, the key link between the quantum formalism and 
experiments, is also involved in the transition from quantum to classical. But decoherence viewed 
from the vantage point of envariance may look unfamiliar. 
 
What other landmarks of decoherence can we get to without using trace and reduced density matrices 
(which rely on Born's rule -- something we do not yet have)? The answer is -- all the essential ones (Zurek, 2005; 2007). We have seen in Section II that pointer states (states that retain correlations, 
and, hence, are predictable and good candidates for classical domain) are singled out directly by the 
nature of information transfers. So we already have a set of preferred pointer states and we have seen that when they are aligned with Schmidt basis, phases between them lose relevance for 
$\cS$ alone. 
Indeed, models of decoherence (Zurek, 1981, 1982, 1991, 2003a;  Paz and Zurek, 2001; Joos et al, 2003, Schlosshauer, 2007) 
predict that after a brief (decoherence time) interlude Schmidt basis will settle down to coincide with 
pointer states determined through other criteria (such as predictability in spite of the coupling to the environment).
 
One more lesson can be drawn from this encounter with decoherence on the way to Born's rule: 
Quantum phases must be rendered irrelevant for additivity of probabilities to replace additivity of 
complex amplitudes. Of course, one could postulate additivity of probabilities by fiat. This was done 
by Gleason (1957), but such an assumption is at odds with the overarching additivity principle of 
quantum mechanics -- with the quantum principle of superposition. So, if we set out with Everett on 
a quest to understand emergence of the classical domain from the quantum domain defined by 
axioms (o) -- (iii), additivity of probabilities should be derived (as it is done in Laplace's approach, 
see Gnedenko, 1968) rather than imposed as an axiom (as it happens in Kolmogorov's measure - theoretic approach, and in Gleason's theorem). 

Assuming decoherence to derive $p_k=|\psi_k|^2$ (Zurek, 1998b; Deutsch, 1999; 
Wallace, 2003) means at best starting  ``half way", and courts circularity (Zeh, 1997; Zurek, 2003a,b; 2005; Schlosshauer, 2007) as physical significance of reduced density matrix -- standard tool of decoherence -- is justified using Born's rule. By contrast, envariant derivation, if successful, can be fundamental, independent of the usual tools of decoherence: It will justify use of 
trace and reduced density matrices in the study of the quantum - classical transition.

There is clearly much more to say about preferred states, einselection, and decoherence, and we shall come back to these subjects later in this section. But now we return to the derivation of Born's rule -- to a problem Everett did appreciate and attempted to solve. 

\subsection{Swaps and equal probabilities}

Envariance of pure states is purely quantum: Classical state of a composite system 
is given by a Cartesian (rather than tensor) product of its constituents. So, to completely 
know a state of a composite classical system one must know a state of each subsystem. 
It follows that when one part of a classical composite system is affected by a transformation
-- a classical analogue of $U_{\cal S}$ -- state of the whole cannot be restored by acting on some 
other part. Hence, {\it pure classical states are never envariant}. 

However, a mixed state (of, say, two coins) can mimic envariance: When we only know that a dime 
and a nickel are `same side up', we can `undo' the effect of the flip of a dime by flipping a nickel. 
This classical analogue depends on a partial ignorance: To emulate envariance, we cannot know individual states of coins, 
just the fact that they are same side up -- just the correlation.

In quantum physics tensor structure of states for composite systems means that
 `pure correlation' is possible. We shall see that a
maximally entangled state with equal absolute values of Schmidt coefficients:
$$|\bar \psi_{\cal SE}\rangle \propto \sum_{k=1}^N e^{-i\phi_k}
|s_k\rangle |\varepsilon_k\rangle  \eqno(3.5)$$ 
{\it implies} equal probabilities for any measurement $\cS$ and $\cE$.
Such an {\it even} state is envariant under a {\it swap} 
$$ u_{\cal S}(k \rightleftharpoons l ) =  |s_k\rangle \langle s_l|  + |s_l\rangle \langle s_k| \  . \eqno(3.6a)
$$
A swap exchanges two cards (Fig.1c). It permutes states $\ket {s_k}$ and $\ket {s_l}$ of the system. 
A swap $\ket {\tt Heads} \bra {\tt Tails} + \ket {\tt Tails} \bra {\tt Heads}$ would flip a coin.

A swap on ${\cal S}$ is envariant when $|a_k| = |a_l|$ because
$u_{\cal S}(k \rightleftharpoons l) $ can be undone by a {\it counterswap} on $\cE$;
$$ u_{\cal E}(k  \rightleftharpoons l) =  e^{i(\phi_k-\phi_l)}|\varepsilon_l\rangle \langle \varepsilon_k| + 
e^{-i(\phi_k-\phi_l)} |\varepsilon_k\rangle \langle \varepsilon_l|  \ .  \eqno(3.6b)$$
Envariance under swaps is illustrated in Fig. 1c.
We want to {\it prove} that probabilities of envariantly swappable outcome states must be equal. 
But let us proceed with caution: Invariance under a swap is not enough -- probability 
could depend on some other `intrinsic' property of the outcome. For instance, in a superposition 
$\ket g +\ket e$, the ground and excited state can be invariantly swapped, but their {\it energies} 
are different. Why shouldn't probability -- like energy -- depend on some intrinsic property of 
the outcome state?

Envariance can be used to prove that it cannot -- that probabilities of envariantly swappable states 
are indeed equal. To prove this, we first define what is meant by ``the state" and ``the system"  
more carefully. Quantum role of these concepts is elucidated by three ``Facts" -- three assumptions 
that are tacitly made anyway, but are worth stating explicitly: 
\begin{description}
\item[{\bf Fact 1}:] Unitary transformations must act on the system to alter
its state. That is, when an operator does not act on the
Hilbert space ${\cal H_S}$ of $\cS$,
i.e., when it has a form $...\otimes {\bf 1}_{\cal S} \otimes...$ 
the state of ${\cal S}$ does not change.
\item[{\bf Fact 2}:] Given the measured observable, the state of the system ${\cal S}$ is all that is needed
(and all that is available) to predict measurement results, including probabilities of outcomes.
\item[{\bf Fact 3}:] The state of a larger composite system that includes
${\cal S}$ as a subsystem is all that is needed (and all that is available)
to determine the state of ${\cal S}$.
\end{description}
\noindent Note that resulting states need not be pure. Also  note that Facts 
-- while  `naturally quantum' -- are not in any obvious conflict with the role 
of states in classical physics. 

We can now {\it prove} that for an even $\ket {\bar \psi_{\cS \cE}}$, Eq. (3.5), state of $\cS$ alone
is invariant under swaps: Swap changes partners in the Schmidt decomposition (and, therefore, 
alters the global state). But, when the coefficients of the swapped outcomes differ only by phase, 
swap can be undone without acting on ${\cS}$ -- by a counterswap in $\cE$.  
As the global state of $\cS \cE$ is restored, it follows (from fact 3) that the state of $\cS$ 
must have been also restored. But, (by fact 1) state of $\cS$ could not have been affected by 
a counterswap acting only on ${\cE}$. So, (by fact 2) the state of $\cS$ must be left 
intact by a swap on $\cS$ corresponding to an even $|\bar \psi_{\cal SE}\rangle$. QED.

We conclude that  envariance of a pure global state of $\cS \cE$ under swaps implies invariance of a corresponding local state of $\cS$. We could now follow Laplace, appeal to indifference, apply equal likelihood, and ``declare victory'' -- claim that subjective probabilities must be equal. However, as we 
have seen with the example of the eigenstates of energy,  invariance of a local state under a swap implies only that probabilities get swapped when outcomes are swaped. This does not yet prove they are equal!

Envariance will allow us to get rid of subjectivity altogether. The simplest way to establish this desired equality is based on perfect correlation between Schmidt states of $\cS$ and $\cE$. These are relative states in the sense of Everett, and they are orthonormal, so they are correlated one-to-one. This implies the same probability for each member of a pair. Moreover (and for the same reason) after a swap 
on $\cS$ probabilities of swapped states must be the same as probabilities of their two new partners in 
$\cE$. But (by Fact 1) the state of $\cE$ (and, by Fact 2, probabilities it implies) are not affected by the swap in $\cS$. So, {\it swapping Schmidt states of} $\cS$ {\it exchanges their probabilities, and when
the entangled state is even it also keeps them the same!} This can be true only if probabilities of envariantly swappable states are equal.
QED. 

We can now state our conclusion:  When all $N$ coefficients in Schmidt decomposition have 
the same absolute value (as in Eq. (3.5)), probability of each Schmidt  state is the same, and, 
by normalization, it is $p_k=1/N$.\footnote{There is an amusing corollary to this theorem: One can now 
{\it prove} that states which appear in a Schmidt decomposition with coefficients $a_l = 0$ have 
0 probability: To this end, consider a decomposition that has $n$ such states. One can combine
two of these states to form a new state, which still has the same coefficient of 0. This purely 
mathematical step that cannot have any physical implications for probabilities of states that were 
not involved. Yet, there are now only $n-1$ states with equal coefficients. So the probability $p_l$ of 
any state with zero amplitude has to satisfy $n p_l=(n-1)p_l$, which holds only when $p_l=0$.} 
Reader may regard this as obvious, but (as noted by Schlosshauer and Fine (2005) 
and Barnum (2003)), this is actually the hard part of the derivation,  as it requires 
establishing a connection between quantum physics and mathematics with only minimal 
set of assumptions at hand. Still, this may seem like a lot of work to establish something `obvious': 
The case of unequal coefficients is our real goal. But -- as we now show -- it can be reduced 
to the equal coefficient case we have just settled. 

It is important to emphasize that in contrast to many other approaches to both classical and quantum probability, our envariant derivation is based not on a subjective assessment of an observer, but on 
an objective, experimentally verifiable symmetry of entangled states. Observer is forced infer equal 
probabilities not because of his ignorance, but because his certainty about something else -- about a global state of the composite system -- implies that local states are completely unknown.

\subsection{Born's rule from envariance}

To illustrate general strategy we start with an example involving
a two-dimensional Hilbert space of the system spanned by states
$\{|0\rangle,|2\rangle\}$ and (at least) a three-dimensional Hilbert space
of the environment:
$$ |\psi_{\cal SE} \rangle \ \propto \ \sqrt{\frac 2 3}~|0\rangle_{\cS}|+ \rangle_{\cE} \ \ + \ \  \sqrt{\frac 1 3} ~|2\rangle_{\cS}|2\rangle_{\cE}  \ .  \eqno(3.7a)$$
System is represented by the leftmost kets, and
$|+\rangle_{\cE}=(|0\rangle_{\cE}+|1\rangle_{\cE})/\sqrt 2$ exists
in (at least two-dimensional) subspace of ${\cal E}$ that is orthogonal to
the state $|2\rangle_{\cE}$, so that
$\langle0|1\rangle=\langle0|2\rangle=\langle1|2\rangle=\langle+|2\rangle=0$.
We already know we can ignore phases in view of their irrelevance for states 
of subsystems, so we omitted them above. 

To reduce this case to an even state we extend $|\psi_{\cal SE}\rangle$
above to a state $|\bar \Psi_{\cal SEC}\rangle$ with equal coefficients by letting $\cE$ act on
an ancilla ${\cal C}$. (By Fact 1, since $\cS$ is not acted upon, so probabilities
we shall infer for is cannot change.) This can be done
by a generalization of controlled-not 
acting between ${\cal E}$ (control) and ${\cal C}$ (target), so that
(in the obvious notation) $|k\rangle|0'\rangle \Rightarrow |k\rangle|k'\rangle$, leading to;
$$ 
\sqrt 2 \ket 0 \ket + \ket {0'} + \ket 2 \ket 2 \ket {0'}
\Longrightarrow $$
$$
\Longrightarrow \sqrt 2 |0\rangle{{|0\rangle|0'\rangle + |1\rangle|1'\rangle} \over \sqrt 2}  + |2\rangle|2\rangle|0'\rangle \eqno(3.8a)
$$
Above, and from now on we skip subscripts: State of ${\cal S}$ will be always listed first,
and state of ${\cal C}$ will be primed. The cancellation of $\sqrt 2$ yields equal coefficient state:
$$ |\bar \Psi_{\cal SCE}\rangle \propto |0,0'\rangle|0\rangle + |0,1'\rangle|1\rangle 
 + |2,2'\rangle|2\rangle  \ . \eqno(3.9a)$$
Note that we have now combined state of $\cS$ and ${\cal C}$ and (in the next step) we shall swap states of  ${\cal SC}$ as if it was a single system.

Clearly, for joint states $|0,0'\rangle,~ |0,1'\rangle$, and $|2,2'\rangle $ of ${\cal S}{\cal C}$ 
this is a Schmidt decomposition of ($\cS \cal C) {\cal E}$. The three orthonormal product states have
coefficients with the same absolute value. So, they can be envariantly swapped. It follows that the
probabilities of these Schmidt states --- $|0\rangle|0'\rangle, \ |0\rangle|1'\rangle,$ and $|2\rangle|2'\rangle$ --
are all equal, so by normalization they are $\frac 1 3$. Moreover,
probability of  state $|2\rangle$ of the system is $ \frac 1 3$. As $\ket 0$ and $\ket 2$ are the only
two outcome states for $\cS$, it also follows that probability of 
$|0\rangle$ must be $\frac 2 3$. Consequently:
$$ p_0 =  \frac 2 3; \ \ p_2= \frac 1 3 \ . \eqno(3.10a) $$
This is Born's rule! 

Note that above we have avoided assuming additivity
of probabilities: $ p_0 = \frac 2 3$ not because it is a sum of two fine-grained alternatives each with 
probability $ \frac 1 3$, but rather because there are only two (mutually exclusive and exhaustive) alternatives for $\cS$; $\ket 0$ and $\ket 2$, and $p_2= \frac 1 3$. So, by normalization, $ p_0 = 1-  \frac 1 3$.

Bypassing appeals to additivity of probabilities is a good idea in interpreting a theory with
another principle of additivity -- quantum superposition principle -- which trumps additivity of probabilities or at least classical intuitive ideas about what should be additive (e.g., in the double slit experiment). Here this conflict is averted: Probabilities of Schmidt states can be added because of the loss of phase coherence that follows directly from envariance as we have established earlier (Zurek, 2005).

Consider now a general case.
For simplicity we focus on entangled state
with only two non-zero coefficients:
$$ |\psi_{\cal SE} \rangle = \alpha |0\rangle |\varepsilon_0 \rangle +
\beta |1 \rangle |\varepsilon_1 \rangle \ , \eqno(3.7b)$$
and assume
$ \alpha = \sqrt {\frac m M}; \ 
\beta = \sqrt {{\frac {M-m} M}} $, with integer $m $, $ M$.
\hocom{When there are no $m$ and $M$ for which Eq. (3.10b) holds exactly, we can still
put upper and lower bounds on $|\alpha|$ and $|\beta|$ by taking a sequence
of increasing $M$ and $m_+=m_-+1$ such that $\sqrt{m_+/M} > |\alpha| 
> \sqrt {m_-/M}$ and, by continuity, recover our conclusions in the
$M \rightarrow \infty$ limit.}

As before, the strategy is to convert a general entangled state
into an even state, and then to 
apply envariance under swaps.
\hocom{-based reasoning that has led to Eq. (9). ``Fine-graining'' 
is a well - known trick, used on similar occassions in the classical 
probability textbooks, but applicable also in the quantum context$^{16,17}$. }
To implement it, we assume ${\cal E}$  has sufficient dimensionality to
allow decomposition of $|\varepsilon_0\rangle$ and $|\varepsilon_1\rangle$ in a different 
orthonormal basis $\{ |e_k\rangle \}$:
$$ |\varepsilon_0\rangle = \sum_{k=1}^m |e_k\rangle/\sqrt m ; \ \ \ \ \ 
|\varepsilon_1\rangle = \sum_{k=m+1}^M |e_k\rangle/\sqrt {M-m} $$

Envariance we need is associated with counterswaps 
of ${\cal E}$ that undo swaps of the joint state of the composite
system ${\cal SC}$. To exhibit it, we let ancilla ${\cal C}$ interact with ${\cal E}$ 
as before, e.g. by employing ${\cal E}$ as a control to carry out
$|e_k\rangle |c_0 \rangle \rightarrow |e_k\rangle |c_k \rangle $,
where $|c_0\rangle$ is the initial state of ${\cal C}$ in some suitable
orthonormal basis $\{|c_k\rangle \}$. Thus;
$$ |\bar \Psi_{\cal SCE}\rangle \propto  \sqrt{m} ~ |0\rangle ~ \sum_{k=1}^m  {{ |c_k\rangle |e_k\rangle } \over \sqrt m} \\  
+ \sqrt {M-m}~|1\rangle \sum_{k=m+1}^M {{|c_k\rangle |e_k\rangle}
\over \sqrt{M-m}} \eqno(3.8b) $$
obtains. This ${\cal CE}$ interaction can happen far from ${\cal S}$, so by Fact 1 it
cannot influence probabilities in ${\cal S}$.
$|\bar \Psi_{\cal SCE}\rangle $ is envariant under swaps of states 
$|s, c_k\rangle$ of the composite ${\cal SC}$ system (where $s$ stands 
for 0 or 1, as needed).
This is even more
apparent after the obvious cancellations;
$$ |\bar \Psi_{\cal SCE}\rangle \propto
\sum_{k=1}^m |0, c_k\rangle |e_k\rangle +
 \sum_{k=m+1}^M |1, c_k\rangle |e_k\rangle . \eqno(3.9b)$$
Hence, $ p_{0,k} = p_{1,k} = \frac 1 M $. So,
probabilities of $|0\rangle $ and $|1\rangle$:
$$ p_0 = {m \over M} = |\alpha|^2; \ \ \
p_1 = {{M-m} \over M} = |\beta|^2 \eqno(3.10b)$$
are given by Born's rule. It arises from the most quantum aspects of the theory -- entanglement and envariance. 

In contrast with other approaches, probabilities in our envariant derivation are a consequence of complementarity, of the incompatibility of purity of entangled state of the whole with purity of the 
states of parts. Born's rule arises in a completely quantum setting, without any {\it a priori} imposition 
of symptoms of classicality that violate spirit of quantum theory. In particular, envariant derivation 
(in contrast to Gleason's successful but unphysical proof and to Everett's unsuccessful attempt) does 
not require additivity as assumption: The strategy that bypasses appeal to additivity used in the simple 
case of Eq. (3.10a) can be generalized (Zurek, 2005). In quantum setting this is an important 
advance. It can be made only because relative phases between Schmidt states are envariant -- because of decoherence. The case of more than two outcomes is straightforward, as is extension by continuity to incommensurate probabilities.

\subsection{Relative frequencies from relative states}

We can now use envariance to deduce relative frequencies.
Consider ${\cal N}$ distinguishable ${\cal SCE}$ triplets, all in the state of
Eq. (3.9). The state of the ensemble is then;
$$|\Upsilon^{\cal N}_{\cal SCE}\rangle = \otimes_{\ell=1}^{\cal N}
|\bar \Psi_{\cal SCE}^{(\ell)}\rangle \eqno(3.11)$$  
We now repeat steps that led to Eqs. (3.10) for the ${\cal SCE}$ triplet,
and think of ${\cal C}$ as a counter, a detector in which
states $|c_1\rangle \dots |c_m\rangle$ record ``0'' 
in ${\cal S}$, while $|c_{m+1}\rangle \dots |c_M\rangle$ record ``1''.
Carrying out tensor product and counting terms with $n$ detections
of ``0'' yields the total
$ \nu_{\cal N}(n) = {{\cal N} \choose n} m^n (M-m)^{{\cal N} - n}$ of fine-grained records -- or of the 
corresponding envariantly swappable Everett branches with histories of detections that differ by a sequence of 0's and 1's, and by the labels assigned to them by the counter, but that have the same 
total of 0's. Normalizing leads to probability of a record with $n$ 0's:
$$ p_{\cal N}(n) = {{\cal N} \choose n} |\alpha|^{2n} |\beta|^{2({\cal N}-n)}
\simeq { 
e^{- { 1 \over 2}\bigr({{n - |\alpha|{\cal N}} \over \sqrt {\cal N} |\alpha \beta|}\bigl)^2} 
\over {\sqrt{2 \pi {\cal N}} |\alpha \beta|}}
\eqno(3.12)$$
Gaussian approximation of a binomial is accurate for large ${\cal N}$: We shall assume ${\cal N}$ is
large not because envariant derivation requires this (we have already obtained Born's rule for ${\cal N}=1$), but because relative frequency approach needs it (von Mises, 1939;  Gnedenko, 1968).

The average number of 0's is, according to Eq. (3.12) $\langle n \rangle = | \alpha|^2 {\cal N}$, as expected, establishing a link between relative frequency of events in a large number of trials and Born's rule. This connection between quantum states and relative frequencies does not rest on either circular and {\it ad hoc} assumptions that relate size of the coefficients in the global state vector to probabilities (e.g., by asserting that probability corresponding to a small enough amplitude is 0 (Geroch 1984);  Buniy, Hsu, \& Zee, 2006)), modifications of quantum theory (Weissman, 1999), 
or on the unphysical infinite limit 
(Hartle, 1968; Farhi, Goldstone, and Guttmann, 1989). Such steps have left past frequentist approaches to Born's rule (including also these of Everett, DeWitt, and Graham) open to a variety of criticisms (Stein, 1984;  Kent, 1990; Squires, 1990; Joos, 2000; Auletta, 2000).

In particular, we avoid problem of two independent measures of probability (number of branches 
{\it and} size of the coefficients) that derailed previous relative state attempts. We simply count the 
number of {\it envariantly swappable} (and, hence {\it provably equivalent}) sequences of potential 
events. This settles the issue of  ``maverick universes'' -- atypical branches with numbers of 
e.g. 0's quite different from the average $\langle n \rangle$. They are there (as they should be) but 
are very improbable. This is 
established through a physically motivated envariance under swaps. So, maverick branches did not 
have to be removed either  ``by assumption'' (DeWitt, 1970; 1971; Graham, 1973; Geroch, 1984) 
or by an equally unphysical ${\cal N}= \infty$. 

\subsection{Summary}

Envariance settles major outstanding problem of relative state interpretation: The origin of Born's rule. 
It can be now established without assumption of the additivity of probabilities (as in Gleason, 1957). 
We have also derived $p_k=|\psi_k|^2$ without relying on tools of decoherence.

Recently there were other attempts to apply Laplacean strategy of invariance under permutations to prove ``indifference''.  This author (Zurek, 1998b) noted that all of the possibilities listed on a diagonal 
of a unit density matrix (e.g., $\sim \kb 0 0 + \kb 1 1$) must be equiprobable, as it is invariant under swaps. This approach can be then extended to the case of unequal coefficients, and leads to Born's rule. 
However, a density matrix is not the right starting point for the derivation: A pure state, prepared by 
the observer in the preceding measurement, is. And to get from such a pure state to a mixed (reduced) density matrix one must ``trace'' -- average over e.g., the environment. Born's rule is involved in the 
averaging, which courts circularity (Zeh, 1997; Zurek, 2003a,b; 2005). 

One could of course try to start with a pure state instead. Deutsch (1999) and his followers 
(Wallace, 2002, 2003; Saunders, 2004) pursue this strategy, couched in terms of decision theory. 
The key is again invariance under permutations. It is indeed there for certain pure states (e.g., 
$\ket 0 + \ket 1$) but not when relative phase is involved. That is, $\ket 0 + \ket 1$ equals 
the post-swap $\ket 1 + \ket 0$, but $\ket 0 + e^{\i \phi}\ket 1 \neq \ket 1 + e^{\i \phi}\ket 0$, and the difference is not the overall phase. Indeed, $\ket 0 + i\ket 1$ is {\it orthogonal} to $i\ket 0 + \ket 1$, so there is no invariance under swaps. In an isolated system this problem cannot be avoided. 
(Envariance of course deals with it very naturally.)
The other problem is selection of events one of which will happen upon measurement -- the choice of
preferred states. These two problems must be settled, either through appeal to decoherence (as in Zurek, 1998b, and in Wallace, 2002; 2003), or by ignoring phases essentially ad hoc (Deutsch, 1999), which then makes readers suspect that some ``Copenhagen'' assumptions are involved. Indeed, Barnum et al. (2000) criticized Deutsch (1999) paper interpreting his approach in the 
``Copenhagen spirit''. And decoherence -- invoked by Wallace (2002) -- employs 
reduced density matrices, and, hence, Born's rule (Zurek, 2003b, 2005; Baker, 2007; Forrester, 2007; 
Schlosshauer, 2007). 

Envariant derivation of Born's rule we have presented is an extension extension of the swap strategy in
(Zurek, 1998b): However, instead of tracing the environment, we incorporated it in the discussion. 
This leads to Born's rule, but also to new appreciation of decoherence.  Pointer states can be inferred directly from dynamics of information transfers as was shown in Section II and, indeed, in the original definition of pointer states (Zurek, 1981). Not everyone is comfortable with envariance (see e.g. Herbut, 2007, for a selection of views). But this is understandable -- interpretation of quantum theory is a field 
rife with controversies.

Envariance is firmly rooted in physics. It is based on symmetries of entanglement.  One may be 
nevertheless concerned about the scope of envariant approach:  $p_k=|\psi_k|^2$ for Schmidt states, 
but  how about measurements with other outcomes? 
The obvious starting point for derivation of probabilities is not 
an entangled state of $\cS$ and $\cE$, but a pure state of $\cS$. And such a state can be 
expressed in {\it any} basis that spans $\cH_{\cS}$. So why entanglement? And why Schmidt states?

Envariance of phases of Schmidt coefficients is closely tied to einselection of pointer states: 
After decoherence has set in, pointer states nearly coincide with Schmidt states. 
Residual misalignment is not going to be a major problem. At most, it might 
cause minor violations of laws obeyed by the classical probability for events defined by 
pointer states. Such violations are intriguing, and perhaps even detectable, but unlikely to matter in the macroscopic setting we usually deal with. To see why, we revisit pointer states -- Schmidt states 
(or einselection - envariance) link in the setting of measurements. 

Observer $\cal O$ uses an (ultimately quantum) apparatus ${\cal A}$, initially in a known state 
$\ket {A_0}$, to entangle with $\cS$, which then decoheres as $\cal A$ is immersed in $\cE$ (Zurek, 1991, 2003a; Joos et al., 2003; Schlosshauer, 2004; 2007). This sequence of interactions leads to:
$\ket {\psi_{\cS}} \ket {A_0} \ket {\varepsilon_0} \Rightarrow
\bigl(\sum_k a_k \ket {s_k} \ket {A_k}\bigr) \ket {\varepsilon_0}  \Rightarrow
\sum_k a_k \ket {s_k} \ket {A_k} \ket {\varepsilon_k}$.
In a properly constructed apparatus pointer states $\ket {A_k}$ are unperturbed 
by ${\cal E}$ while $\ket {\varepsilon_k}$ become orthonormal on a decoherence timescale.
So in the end we have Schmidt decomposition of ${\cal SA}$ (treated as a single entity) and ${\cal E}$. 

Apparatus is built to measure a specific observable 
$ {\sigma}_{\cS} = \sum_k \varsigma_k |s_k\rangle \langle s_k|$, 
and $\cal O$ knows that $\cS$ starts in $\ket {\psi_{\cS} } = \sum_k a_k \ket {s_k}$. 
The choice of ${\cal A}$ (of Hamiltonians, etc.) commits observer to a definite set 
of potential outcomes: Probabilities will refer to $\{\ket {A_k }\}$, or, equivalently, 
to $\{\ket {A_k s_k}\}$ in the Schmidt decomposition. So, to answer questions we started with, 
entanglement is inevitable, and only pointer states (e.g., nearly Schmidt states after 
decoherence) {\it of the apparatus} can be outcomes. This emphasis on the role of apparatus  
in deciding what happens parallels Bohr's view captured by  ``No phenomenon is a phenomenon untill it is a recorded phenomenon" (Wheeler, 1983).
However, in our case $\cal A$ is quantum and symptoms of classicality -- e.g., einselection as well
as loss of phase coherence between pointer states -- arise as a result of entanglement with $\cE$.

Envariant approach applies even when $\ket {s_k}$ aren't  orthogonal: 
Orthogonality of $\ket {A_k s_k}$ is assured by $\langle A_k | A_l \rangle =\delta_{kl}$ 
-- by distinguishability of records in a good apparatus. This is because events
that we have direct access to are records in $\cA$ (rather than states of $\cS$). 

Other simplifying assumptions we invoked can be also relaxed (Zurek, 2005). For example, 
when $\cE$  is initially mixed (as will be generally the case), one can `purify'  it by adding extra 
$\tilde \cE$ in the usual manner (see Section II). Given that we already have a derivation of Born's
rule, its use (when justified by the physical context) does not require apologies, and does not introduce
circularity. Indeed, it is interesting to enquire what instances of probabilities in physics {\it cannot} be interpreted envariantly.

Purifications, use of ancillae, fine - graining, and other steps in the derivation
need not be carried out in the laboratory each time probabilities are extracted
from a state vector: Once established, Born's rule is a {\it law}. It follows from the 
geometry of Hilbert spaces for composite quantum systems. We used assumptions
about ${\cal C}, \cE, $ etc., to demonstrate $p_k=|\psi_k|^2$,  but this rule must be obeyed even
when no one is there to immediately verify compliance. So, even when there is no ancilla 
at hand, or when $\cE$ is initially mixed or too small for fine-graining, one could (at some 
later time, using purification, extra environments and ancillae) verify that bookkeeping implicit in
assigning probabilities to $\ket {\psi_{\cS}}$ or pre-entangled $\ket {\psi_{\cS \cE}}$ abides by
the symmetries of entanglement.

The obvious next  question is  how to verify envariance directly. 
One way is to carry out experimentally steps we outlined earlier, Eqs. (3.4)-(3.10). 
To this end, one can e.g. attach ancilla ${\cal C}$, carry out swaps, and make intermediate 
measurements that verify equal probabilities in the fine-grained states of $\cS {\cal C}$ through
swaps. This leads to probability of potential outcomes -- it can be deduced from sequences of 
reversible operations that involve just a single copy of $\ket {\psi_{\cS}}$ (rather than a whole 
ensemble) by determining fraction of equiprobable alternatives $\ket {s_k c_k}$ corresponding
to specific $\ket {s_k}$.
Such sequences of swaps, confirmatory measurements, and counterswaps can be
easily devised on paper, but harder to implement: For instance,
it would be best if measurements probed an entangled {\it global} state
of ${\cal SCE}$, as global outcomes can be predicted with certainty (so that confirming
envariance at the root of Born's rule does not rely on Born's rule). However, even
measurements that destroy global state and require statistics to verify
envariance (e.g., quantum tomography) would be valuable: What is at stake is a basic symmetry 
of nature that provides a key link between the unitary `bare quantum theory' and experiments.

Probabilities described by Born's rule quantify ignorance of $\cal O$ {\it before} he measures. 
So they admit ignorance interpretation -- $\cal O$ is ignorant of the {\it future} outcome, 
(rather than of an unknown {\it pre-existing} real state, as was the case 
classically). Of course, once $\cal O$'s memory becomes correlated with $\cal A$, 
its state registers what $\cal O$ has perceived (say, $\ket {o_7}$ that registers $\ket {A_7}$). 
Re-checking of the apparatus will confirm it. 
And, when many systems are prepared in the same initial state, 
frequencies of outcomes will be in accord with Born's rule. 

Envariant approach uses incompatibility between observables of the whole and its parts. In retrospect 
it seems surprising that envariace was not noticed and used before to derive probability or to provide 
insights into decoherence and environment - induced superselection: Entangling interactions are key 
to measurements and decoherence, so entanglement symmetries would seem relevant. However, entanglement is often viewed as paradox, as something that needs to be explained, and 
not used in an explanation. This attitude is, fortunately, changing.

\section{Quantum Darwinism}

Objective existence in quantum theory is a consequence of a {\it relationship} between the system 
and the observer, and not just (as was the case classically) ``sole responsibility'' of the system. 
This relational view of existence is very much in concert with Everett: Relative states can exist 
objectively, providing that observer will only measure observables that commute with the preexisting mixed state of the system (e.g., in the wake of decoherence, its pointer observable). But why should 
the observers measure only pointer observables?

Quantum Darwinism provides a simple and natural explanation of this restriction, and, hence, of the 
objective existence -- bulwark of classicality -- for the einselected states: Information we acquire about 
the ``rest of the Universe'' comes to us indirectly, through the evidence systems of interest deposit in their 
environments. Observers access directly only the record made in the environment, an 
imprint of the original state of $\cS$ on the state of a fragment of $\cE$. And there are multiple copies of that original (e.g., of this text) that are disseminated by the photon environment, by the light that is 
either scattered (or emitted) by the printed page (or by the computer screen). We can find out the state
of various systems indirectly, because their correlations with $\cE$ (which we shall quantify below 
using mutual information) allow $\cE$ to be a witness to the state of the system.

\begin{figure}[tb]
\begin{tabular}{l}
\vspace{-0.15in} \includegraphics[width=\FCW]{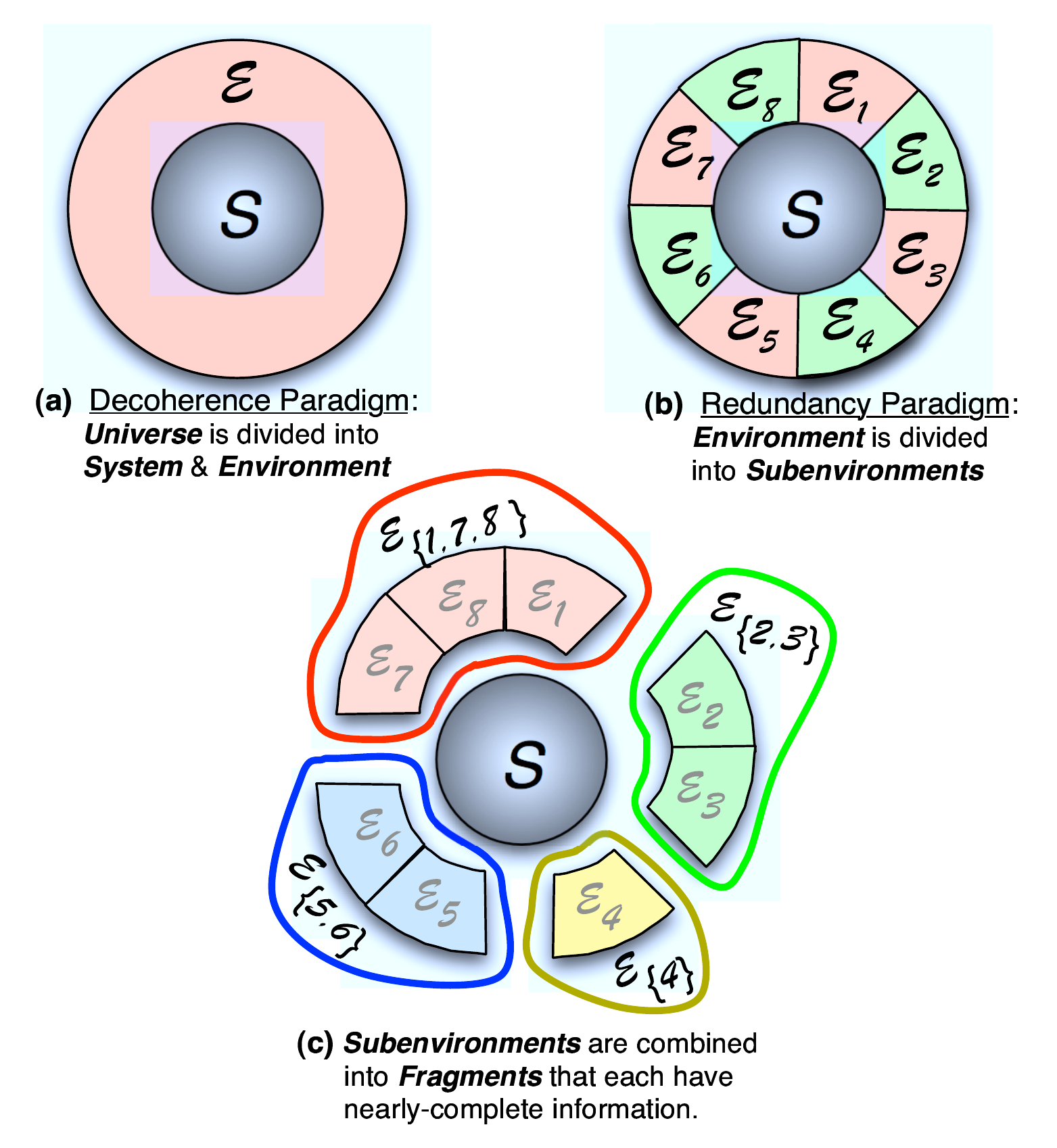}\\
\end{tabular}
\caption{\emph{Quantum Darwinism and the structure of the environment}.  The decoherence paradigm distinguishes between a system ($\Sys$) and its environment ($\Env$) as in \textbf{(a)}, but makes no further recognition of the structure of $\cE$; it could be as well monolithic.  In the environment-as-a-witness paradigm, we recognize subdivision of $\Env$ into subenvironments -- its natural subsystems, as in \textbf{(b)}. The only real requirement for a subsystem is that it should be individually accessible to measurements; observables corresponding to different subenvironments commute. To obtain information about the system $\cS$ from its environment $\cE$ one can then carry out measurements 
of \emph{fragments} $\cF$ of the environment -- non-overlapping collections of the subsystems of $\cE$.  Sufficiently large fragments of $\cE$ that has monitored (and, therefore, decohered) $\cS$ can often 
provide enough information to infer the state of $\Sys$, by combining subenvironments as in \textbf{(c)}.  
There are then many copies of the information about $\cS$ in $\cE$, which implies that some information 
about the ``fittest'' observable that survived monitoring by $\cE$ has proliferated throughout $\cE$, leaving multiple informational offspring. This proliferation of the information about the fittest states defines
quantum Darwinism. Multiple copies allow many observers to find out the state of $\cS$: Environment becomes a reliable witness with redundant copies of information about preferred
observables, which leads to objective existence of preferred pointer states.}
\label{EnvSubdivision}
\end{figure}

The plan of this section is to define mutual information, and to use it to quantify
information that can be gained about $\cS$ from $\cE$. Objectivity arises because the same information
can be obtained independently by many observers from many fragments of $\cE$. So, to
quantify objectivity of an observable we compute its redundancy in $\cE$ -- count the 
number of copies of its record. 

Results of two previous sections will be useful: Derivation of the pointer observable we have
reported in Section II allows us to anticipate what is the preferred observable capable of leaving
multiple records in $\cE$: Only states that can be monitored without getting perturbed can survive 
long enough to deposit multiple copies of their information - theoretic progeny in the environment. 
In other words, the no-cloning theorem that is behind the derivation of Section II is not an obstacle 
when ``cloning'' involves not an unknown quantum state, but rather an einselected observable that is 
``fittest'' -- that has already survived evolutionary pressures of its environment and can produce 
copious information - theoretic offspring.

In order to make these arguments rigorous we shall compute entropies of $\cS$, $\cE$, and various 
fragments $\cF$ of $\cE$ (see Fig. 2). This means we will need reduced density matrices and probabilities. It is
therefore fortunate that, in Section III, we have arrived at a fundamental derivation of Born's rule from
symmetry of entanglement. This gives us a right to use the usual tools of decoherence -- 
reduced density matrix and trace -- to compute entropy. 

\subsection{Mutual Information, redundancy, and discord}

Quantities that play key role in quantum Darwinism
are often expressed in terms of von Neumann entropy:
$$ H(\rho) = - Tr \rho \lg \rho \  , \eqno(4.1) $$
The density matrix describes the state of a system, or of a collection of several quantum systems. As is the case classically, von Neumann entropy in the quantum case is a measure 
of ignorance. But density matrix $\rho$ provides more that just its eigenvalues that determine $H(\rho)$ -- more than just the set of probabilities. It is an operator -- it has eigenstates. So, one is tempted to add 
that $\rho$ also determines what one is ignorant of. This is not necessarily the case: As Section II has 
demonstrated, pointer states do not in general coincide with the states on the diagonal of the reduced density matrix. Therefore, the set of states that one should be curious about (because of their stability) 
may not coincide with the instantaneous eigenstates of the reduced density matrix (although 
after decoherence time they should be approximately aligned).

Nevertheless, one might be interested in the information about some other observable that has
eigenstates $\{ \ket {\pi_k} \}$ which differ from the eigenstates of $\rho$. The corresponding entropy is
then Shannon entropy given by:
$$ H(p_k) = - \sum_k p_k \lg p_k \ , \eqno(4.2)$$
where:
$$ p_k = Tr \bra {\pi_k} \rho \ket {\pi_k} \eqno(4.3)$$
are the associated probabilities. For example, $\{ \ket {\pi_k} \}$ may be the pointer states. They will 
coincide with the eigenstates of the reduced density matrix $\rho_\cS$ only after ``decoherence has 
done its job''. In this case, the difference between the two entropies disappears.

Strictly speaking, it is only then that one can associate the usual interpretation of probabilities to the 
pointer states. The problem one can encounter when this is not so is apparent in composite 
quantum systems. We shall exhibit it by computing mutual information  which characterizes 
correlations between systems. Mutual information will also be our measure of how much (and what)
fragment of the environment knows about the system. 

\subsubsection{Mutual information}

Mutual information is the difference between the entropy of two systems treated separately and jointly:
$$ I(\cS : \cA) = H(\cS)+ H(\cA) - H(\cS, \cA) \ . \eqno(4.4) $$
For classical systems this definition is equivalent to the definition that employs conditional information
(e.g., $H(\cS|\cA)$). It is defined by separating out of the joint entropy $H(\cS, \cA)$ 
the information about one of the two systems. For example:
$$ H(\cS, \cA)=H(\cA)+H(\cS|\cA) = H(\cS) + H(\cA | \cS) \  . \eqno(4.5)$$
Conditional entropy quantifies information about $\cS$ (or $\cA$) that is still missing even after
the state of $\cA$ (or $\cS$) is already known. 

In quantum physics ``knowing'' is not as innocent as in the classical setting: It involves performing a measurement, which in turn alters the joint density matrix into the outcome - dependent {\it conditional
density matrix} that describes state of the system given the measurement outcome -- e.g., state
$\ket {A_k}$ of the apparatus $\cA$:
$$ \rho_{\cS \ket {A_k}} = \bra {A_k} \rho_{\cS\cA} \ket {A_k} / p_k \ , \eqno(4.6)$$
where, in accord with Eq. (4.3), $p_k = Tr  \bra {A_k} \rho_{\cS\cA} \ket {A_k}$.
Conditional entropy given an outcome $ \ket {A_k}$ is then:
$$ H (\cS \ket {A_k}) = -Tr \rho_{\cS \ket {A_k}} \lg  \rho_{\cS \ket {A_k}} \ , \eqno(4.7)$$
which leads to the average:
$$ H (\cS | \{ \ket {A_k}\}) = \sum_k p_k H (\cS \ket {A_k}) \ . \eqno(4.8)$$
So this is how much information about $\cS$ one can expect will be still missing after the measurement
of the observable with the eigenstates $\{ \ket {A_k}\}$ on $\cA$. 

One can pose a more specific question -- e.g., how much information about a specific observable 
of $\cS$ (characterized by its eigenstates $\{ \ket {s_j} \}$) will be still missing after the observable with 
the eigenstates $\{ \ket {A_k}\}$ on $\cA$ is measured. This can be answered by using $\rho_{\cS\cA}$
to compute the joint probability distribution:
$$ p(s_j, A_k) = \bra {{s_j},  {A_k}} \rho_{\cS\cA} \ket {{s_j},  {A_k}} \ . \eqno(4.9)$$
These joint probabilities are in effect classical. They can be used to compute classical joint entropy 
for any two observables (one in $\cS$, the other in $\cA$), as well as entropy of each 
of these observables separately, and to obtain the corresponding mutual information:
$$ I(\{ \ket {s_j} \}:\{ \ket {A_k}\}) = H(\{ \ket {s_j} \}) + H(\{ \ket {A_k}\}) - H(\{ \ket {s_j} \}, \{ \ket {A_k}\}) \ . \eqno(4.10)$$
We shall find uses for both of these definitions of mutual information. In effect, the von Neumann entropy
based $I(\cS:\cA)$, Eq. (4.4), answers the question ``how much the systems know about each other'', 
while the Shannon version immediately above quantifies the mutual information between two specific observables. Shannon version is (by definition) basis dependent. It is straightforward to see (extending
arguments of Ollivier and Zurek (2002)) that, for the same underlying joint density matrix:
$$ I(\cS : \cA) \geq I(\{ \ket {s_j} \}:\{ \ket {A_k}\})  \ . \eqno(4.11)$$
Equality can be achieved only for a special choice of the measured observables, and only when the
eigenstates of $\rho_{\cS\cA}$ are not entangled.

One can also define ``half way'' (Shannon - von Neunann) mutual informations which presume a specific measurement on one of the two systems (e.g., $\cA$), but make no such commitment for 
the other one. For instance, 
$$J(\cS:\{ \ket {A_k}\})=H(\cS)- H (\cS | \{ \ket {A_k}\}) \ \eqno(4.12)$$ 
would be one way to express the mutual information defined ``asymmetrically" in this manner. There 
are some subtleties involved in such definition (Zurek, 2003c), and we shall not pursue this asymmetric 
subject in much greater detail because we can discuss quantum Darwinism without making extensive 
use of $ J(\cS:\{ \ket {A_k}\})$ and similar quantities. Nevertheless, $J(\cS:\{ \ket {A_k}\})$ can be used
to quantify ``quantumness'' of correlations.

{\it Quantum discord} is a difference between mutual entropy defined using the symmetric 
von Neumann formula, Eq. (4.4), and one of the Shannon versions. For example:
$$\delta I (\cS:\{ \ket {A_k}\})= I(\cS : \cA) - J(\cS:\{ \ket {A_k}\}) \ . \eqno(4.13)$$
Discord defined in this manner is a measure of how much information about the two systems is
accessible through a measurement on $\cA$ with outcomes $\{ \ket {A_k}\}$. It clearly basis-dependent,
and the minimum discord disappears:
$$\delta I (\cS,\cA)=min_{\{ \ket {A_k}\}} \{ \delta I (\cS:\{ \ket {A_k}\}) \} =0 \ \eqno(4.14)$$
iff $\rho_{\cS\cA}$ commutes with ${\bf A} = \sum_k \alpha_k \kb  {A_k}  {A_k} $ (Ollivier and Zurek, 2002).
When that happens, quantum correlation is classically accessible from $\cA$. It is of course possible to
have correlations that are accessible only ``from one end'' (Zurek, 2003c). For instance:
$$\rho_{\cS\cA}=\frac 1 2 (\kb \uparrow \uparrow \kb {A_\uparrow} {A_\uparrow} + \kb \nearrow \nearrow \kb {A_\nearrow} {A_\nearrow}) \  \eqno(4.15)$$
will be classically accessible through a measurement with orthogonal records
$\{\ket {A_\uparrow},\ket {A_\nearrow}\}$ on $\cA$, but classically inaccessible to any measurement 
on $\cS$ when $\bk  \uparrow \nearrow \neq 0$.

Questions we shall analyze using mutual information will concern both; (i) redundancy of the information 
(e.g., how many copies of the record does the environment have about $\cS$), and; (ii) what is this information about (that is, what observable of the system is recorded in the environment with largest
redundancy). 

One might be concerned that having different measures -- different mutual informations -- could be 
a problem, as this could lead to different answers, but in practice this never becomes 
a serious issue for two related reasons: There is usually a well - defined pointer observable that 
obviously minimizes discord, so various possible definitions of mutual information tend to coincide 
where it matters. Moreover, the effect we are investigating -- quantum Darwinism -- is not subtle: We shall see there are usually
many copies of pointer states of $\cS$ in $\cE$, and the discrepancy between redundancies 
computed using different formulae for mutual information -- differences between numbers of copies defined through different measures -- is dwarfed by that redundancy. 

In other words, questions that are of interest are for example ``What observable of the system can be
inferred from a fragment of $\cE$ when redundancy is large?'' rather than ``What precisely is the 
redundancy of this observable?''. We are investigating emergence of classical properties -- objectivity 
that appears in the limit of large redundancy. There, the precise value of redundancy has as little physical significance as the precise number of atoms on thermodynamic properties of a large system. 

\subsubsection{Evidence and its Redundancy}

We shall study a system $\cS$ interacting with a composite environment $\cE=\cE_1\otimes\cE_2\otimes\dots\otimes\cE_{\cN}$. The question we shall consider concerns the information one can obtain about 
$\cS$ from a fraction $\cF$ of the environemnt $\cE$ consisting of several of its subsystems (see Fig. 2). 
To be more specific, we partition $\cE$ into non-overlapping fragments $\cF_k$. Redundancy of 
the record is then defined as the number of disjoint fragments each of which can supply sufficiently 
complete (e.g., all but a fraction $\delta$) information about $\cS$.

The first question we need to address is:  ``How much information about $\cS$ can one
get from a typical fragment $\cF$ of $\cE$ that contains a fraction 
$$f=
\frac {\#~ of \  subsystems \  in \ \cF} {\# ~of \ subsystems \  in \  \cE} \eqno(4.16)$$ 
of $\cE$?". Or, to put it slightly differently, we ask about the dependence of the mutual 
information $I(\cS, \cF_f)$ on $f$. 

\begin{figure}[tb]
\begin{tabular}{l}
\vspace{-0.15in} \includegraphics[width=\FCW]{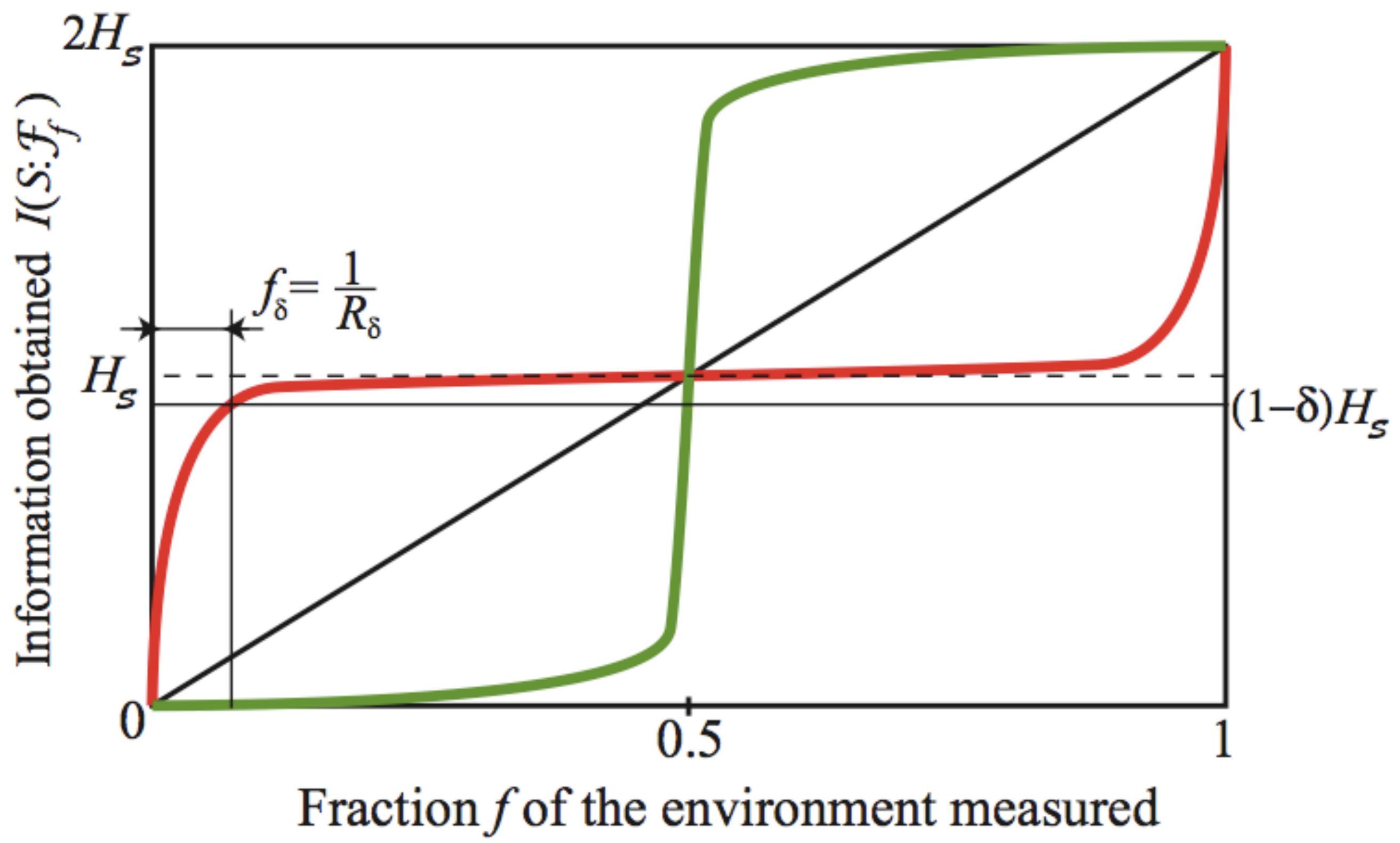}\\
\end{tabular}
\caption{\emph{Partial Information Plot (PIP) and redundancy $R_\delta$ of the information 
about $\cS$ stored in $\cE$}.  When global state of $\cS\cE$ is pure, mutual information that can be 
attributed to a fraction $f$ of the environment must be antisymmetric around the point marked by half 
(50\%) of the environment, and it must be monotonic in $f$. For pure states picked out at random from 
the combined Hilbert space $\cH_{\cS\cE}$, there is very little mutual information between $\cS$ and 
a typical fragment $\cF$ smaller than about half of $\cE$. However, once threshold fraction $\frac 1 2$ 
is attained, nearly all information is in principle at hand. Thus, such random states (green line above)
exhibit no redundancy. By contrast, states of $\cS\cE$ created by decoherence (where the environment $\cE$ monitors preferred observables of $\cS$) allow one to gain almost all (all but $\delta$) of the information about $\cS$ accessible through local measurements from a small fraction $f_\delta=1/R_\delta$ of $\cE$. The corresponding PIP (red line above) quickly asymptotes to $H_\cS$ -- entropy 
of $\cS$ due to decoherence -- which is all of the information about $\cS$ available from measurements
on either $\cE$ of $\cS$. (More information can be ascertained only through global measurements on 
$\cS$ and a fragment $\cF$ corresponding to more than half of $\cE$). $H_\cS$ is therefore the 
\emph{classically accessible information}. As $(1-\delta)H_\cS$ of information can be obtained from 
a fraction $f_\delta=1/R_\delta$ of $\cE$, there are $R_\delta$ such fragments in $\cE$, and $R_\delta$ 
is the {\it redundancy} of the information about $\cS$. Large redundancy implies objectivity: The state 
of the system can be found out indirectly and independently by many observers, who will agree about 
their conclusions. In contrast to direct measurements, it will not be perturbed in this indirect process. Thus, \emph{quantum Darwinism accounts for the emergence of objective existence}.}
\label{SnapPIP}
\end{figure}

This can be illustrated by the partial information plot, or ``PIP" (see Fig. 3) of the von Neumann 
mutual information. The first observation is that these plots are asymmetric around $f=\frac 1 2$.
This can be demonstrated assuming (i) $\cF$ is ``typical'', (ii) the whole of $\cS\cE$ pure, and using elementary properties of the von Neumann mutual information (Blume-Kohout and Zurek, 2005).  

There is a significant difference between the character of PIPs for random pure states in the whole 
joint Hilbert space $\cH_{\cS\cE}=\cH_{\cS}\otimes\cH_{\cE}$ and states resulting from decoherence 
- like evolution: For a random state very little information obtains from fragments with $f < \frac 1 2$. 
By contrast, for PIPs that result from decoherence even a small fragment will typically supply nearly 
all information that can be obtained from less than almost all $\cS\cE$. 

The character of these decoherence - generated PIPs suggest dividing information into easily 
accessible {\it classical  information} $\sim H_\cS$ that can be inferred from any sufficiently large fragment $\cF$ which is still small compared to half of $\cE$, and quantum information 
accessible only through a global measurement on $\cS\cE$. 

This shape of PIPs is a result of einselection: When there is a preferred observable in $\cS$ that
is monitored but not perturbed by the environment, the information about it is recorded over
and over again by different subsystems of $\cE$. So it comes as no surprise that in the end nearly all 
of the easily accessible information can be recovered from any small fragment of the environment.

In this setting $\cF$ plays the role of an apparatus designed to access the same pointer observable 
that can survive intact in spite of the immersion of $\cS$ in $\cE$. This leads to a simplification that
stems from decoherence which singles out the preferred observable: After decoherence sets in, 
density matrix of the system is diagonal in the Schmidt basis which -- by then -- aligns with the pointer basis defined by its resilience. Consequently, for any $\cF$ that is small enough
to let the rest of the environment $\cE_{/\cF}$ (``$\cE$ less $\cF$'') keep $\cS$ decohered, joint density matrix:
$$ \rho_{\cS\cF} = \sum_k p_k \kb {\sigma_k} {\sigma_k} \kb {\cF_k} {\cF_k}=Tr_{\cE_{/\cF}}\rho_{\cS\cE}\   \eqno(4.17)$$
commutes with;
$$ \rho_{\cS} = \sum_k p_k \kb {\sigma_k} {\sigma_k} \  ,  \eqno(4.18)$$
where $\ket {\sigma_k}$ are orthogonal pointer states.

Clearly, $H(\cS)=H(\cS,\cF)= - \sum_k p_k \lg p_k$, even when the states of the environment fragments 
$\{ \ket {\cF_k}\}$ are not orthogonal. Consequently:
$$ I(\cS : \cF) = H(\cS)+ H(\cF) - H(\cS, \cF) = H(\cF) \ . \eqno(4.19) $$
in this case of effective decoherence. Moreover,
$$  I(\{\ket {\sigma_k}\}  : \cF) = I(\cS : \cF) = H(\cF) \ , \eqno(4.20) $$
so that all of the entropy $H(\cF)$ is due to correlations with $\cS$ -- it is the mutual information 
$\cF$ has about $\cS$. 

So, in the limit of effective decoherence the density matrix of the fragment $\cF$ 
$$ \rho_\cF=\sum_k p_k \kb {\cF_k} {\cF_k} \eqno(4.21)$$
holds answers to the remaining questions. Of course, $H(\cF)$ is generally {\it less} than 
$H(\cS)=H(\cS,\cF)= - \sum_k p_k \lg p_k$, as states $\{ \ket {\cF_k}\}$ of the environment fragment  
correlated with pointer states $\{ \ket {\sigma_k}\}$ are in general not orthogonal, so $\{ p_k \}$
are not the eigenvalues of $ \rho_\cF$.

There are two obvious and physically realistic situations that lead to 
$\bk {\cF_k} {\cF_l} \approx \delta_{kl}$, and, hence, to $H(\cF)$ approaching 
$H(\cS)=H(\{\ket {\sigma_k}\})= - \sum_k p_k \lg p_k$. The first one is evolution that leads 
to ever closer correlation of $\cS$ and the same fragment $\cF$: In principle, any fragment $\cF$
large enough  ($Dim \cH_\cF \geq Dim H_\cS$) can hold all of the information about $\cS$, 
and, hence, acquire orthogonal records, $\bk {\cF_k} {\cF_l} = \delta_{kl}$. However, even at 
a fixed time, or even when the evolution has reached a steady state and correlations with $\cS$ 
are no longer increasing, one can gain more information about $\cS$ by intercepting a bigger 
fraction of $\cE$: As the size of the fragment $\cF$ increases, states $\{ \ket {\cF_k}\}$ will hold 
a better record of the pointer observable with which they are correlated: 
They will be more mutually orthogonal. 

There is an interesting and useful corollary to the above discussion: Consider split of the whole composite $\cS\cE$ that separates $\cF$ from the rest (that we shall designate by $\cS\cE_{/\cF}$,
or ``the system $\cS$ and all of the environment $\cE$ except for the fragment $\cF$''). The state vector
of the whole $\cS\cE$ can be written as:
$$ \ket {\Psi_{\cS\cE}}= \sum_k e^{\i \phi_k} \sqrt {p_k} \ket {\cF_k} \ket {\sigma_k} \ket {\cE_{/\cF}|k} \ . \eqno(4.22)$$ 
When (as we have obviously assumed) the whole $\cS\cE$ is in a pure state, the entropy of $\cF$ must 
be the same as the entropy of $\cS\cE_{/\cF}$. Moreover, $\{ \ket {\sigma_k} \ket {\cE_{/\cF}|k}\}$ are orthogonal, and so are $\{ \ket {\sigma_k}\}$. Therefore, the entropy of $\cF$ is the same as the entropy of $\cS$ decohered only by $\cF$:
$$ \rho_{\cS d \cF} = \sum_{k,l} e^{\i (\phi_l-\phi_k)} \sqrt{{p_k}{p_l}} \bk {\cF_k} {\cF_l}  \kb {\sigma_k} {\sigma_l}  \ . \eqno(4.23)$$
This is a useful observation. Let us restate it for the record: {\it When $\cS\cE$ is in a pure state, entropy of a fragment $\cF$ is equal to $H(\cS d \cF)$ -- entropy the system $\cS$ would have 
if it got decohered by $\cF$ alone}. Moreover, in the limit of effective decoherence $I(\cS:\cF)=H(\cF)=H(\cS d \cF)$. The utility of this result has to do with the fact that now one can use decoherence theory 
(albeit in the unfamiliar case of relatively small environments $\cF$) to calculate density matrix of a 
system as it is decohered by $\cF$. The entropy $H(\cS d \cF)$ of $\rho_{\cS d \cF}$ is how much 
$\cF$ ``knows'' about $\cS$.

Convergence of $H_\cF$ (and, hence, of $I(\cS:\cF)$) to $H_\cS$ with the increase in the fragment size 
we exhibited above hints at redundancy:  When $H_\cF \approx H_\cS$ is reached already 
for a typical fragment that is a small fraction $f$ of all the environment, then there are many 
($1/f$) such fragments that can independently provide the same information about $\cS$. Indeed, we 
have arrived -- to within one additional refinement -- at the definition of redundancy: We define 
redundancy as the number of fragments that can independently supply all but $\delta$ of the missing 
information about the system:
$$ R_\delta = \frac 1 f_\delta \ , \eqno(4.24)$$
where $1>\delta \geq 0 $ is defined by:
$$ I(\cS:\cF)= (1-\delta)H(\cS) \ . \eqno(4.25)$$
This definition of redundancy can be illustrated graphically using PIP's: In effect, redundancy is the length of the plateau measured in units set by the support of the initial portion of the graph, the
part starting at $I(\cS:\cF)=0$ and ending when $I(\cS:\cF)=(1-\delta)H(\cS)$ (see Fig. 3).
The reason we have introduced $\delta$ is obvious: $I(\cS:\cF)$ can reach $H(\cS)$ only when 
$f= \frac 1 2$. Thus, large redundancy can be attained only when we relax requirements
on the completeness of information about $\cS$.

\subsection{Quantum Darwinism in Action}

Dissemination of information through the environment has not been analyzed until recently. 
Given the complexity of this process, it is no surprise that the number of results to date
is still rather small. We start with an overview of general features of quantum Darwinism 
we shall then illustrate on specific models. 

Thus, first and foremost; (i) {\it dynamics responsible for decoherence is capable of imprinting multiple 
copies of the information about the system in the environment}. Whether environment can serve 
as a useful witness depends on the room it has to store this information, and whether 
it is stored in places accessible to observers. In other words, quantum Darwinism
will always lead to decoherence, but the reverse is not true: There are situations where environment
cannot store any information about $\cS$. So, redundancy is not implied by decoherence. Moreover; (ii) {\it redundancy can keep on increasing 
long after decoherence has rendered a perfectly einselected} $\rho_{\cS}$. Thus, the number of copies
can continue to grow after the system has decohered. 
Last not least; (iii) {\it only the einselected pointer observable (selected by its predictability) can be redundantly recorded in $\cE$.} While multiple copies of information about the preferred 
observable are disseminated throughout $\cE$, only one copy of the complementary information
is (at best) shared by all the subsystems of the environment, making it in effect inaccessible.

\subsubsection{{\tt c-nots} and qubits}

The simplest model of quantum Darwinism is a rather contrived arrangement of many ($N$) target 
qubits that constitute subsystems of the environment interacting via a {\it controlled not} (``{\tt c-not}'') with 
a single control qubit $\cS$. As time goes on, consecutive target qubits become imprinted with 
the state of the control $\cS$:
$$(a\ket 0+ b\ket 1)\otimes \ket {0_{\varepsilon_1}} \otimes \ket {0_{\varepsilon_2}}\dots \otimes \ket {0_{\varepsilon_N}} \Longrightarrow $$
$$
(a\ket 0\otimes \ket {0_{\varepsilon_1}} \otimes \ket {0_{\varepsilon_2}} + b\ket 1\otimes \ket {1_{\varepsilon_1}} \otimes \ket {1_{\varepsilon_2}})\dots \otimes \ket {0_{\varepsilon_N}} \Longrightarrow 
$$
$$
a\ket 0\otimes \ket {0_{\varepsilon_1}} \otimes \dots \otimes \ket {0_{\varepsilon_N}} + b\ket 1\otimes \ket {1_{\varepsilon_1}} \dots \otimes \ket {1_{\varepsilon_N}} \ . \eqno(4.26)$$
It is evident that this dynamics is creating multiple records of the logical basis states of the system 
in the environment. Mutual entropy between $\cS$ and a subsystem $\cE_k$ can be easily computed. As the $k$'th {\tt c-not} is carried out, $I(\cS : \cE_k)$ increases from 0 to:
$$ I(\cS : \cE_k)=H(\cS)+ H(\cE_k) - H(\cS, \cE_k)=|a|^2\lg|a|^2+|b|^2\lg|b|^2 \  \eqno(4.27)$$ 
Thus, each $\cE_k$ constitutes a sufficiently large fragment 
of $\cE$ to supply complete information about the pointer observable of $\cS$. The very first {\tt c-not}
causes complete decoherence of $\cS$ in its pointer basis $\{ \ket 0, \ket 1\}$. This illustrates points (i) - (iii) above -- the relation between decoherence and quantum Darwinism, 
the continued increase of redundancy well after coherence between pointer states was lost, and the
special role of the pointer observable.

As each environment qubit is a perfect copy of $\cS$, redundancy in this simple example is eventually
given by the number of fragments -- that is, in this case by the number of the environment qubits -- that
have a complete information about $\cS$, e.g. $R=N$. There is no reason to define redundancy 
in a more sophisticated manner, using $\delta$: Such need will arise only in the more realistic cases 
when the analogues of {\tt c-not}'s are imperfect.

Partial information plots in our example would be trivial: $I(\cS:\cF)$ jumps from $0$ to the ``classical''
value given by $H(\cS)=|a|^2\lg|a|^2+|b|^2\lg|b|^2$ at $f=1/N$, continues along the plateau at that level
until $f=1-1/N$, and eventually jumps up again to twice the classical level as the last qubit is included:
The whole $\cS\cE$ is still in a pure state, so when $\cF = \cE$, $H(\cS, \cF)=0$. However, this much information is in a global entangled states, and is therefore accessible only through global measurements.

Preferred pointer basis of the control $\cS$ is of course its logical basis $\{ \ket 0, \ket 1 \}$. These pointer
states are selected by the ``construction'' of {\tt c-not}'s. They remain untouched by copying into consecutive environment subsystems $\cE_k$. As we have already anticipated, after the decoherence
takes place;
$$ I(\cS:\cF)=I(\{ \ket 0, \ket 1 \}:\cF) \ . \eqno(4.28)$$
for any fragment of the environment when there is at least one subsystems of $\cE$ correlated with
$\cS$ left outside of $\cF$, which suffices to decohere $\cS$. When this is the case, minimum quantum discord disappears: 
$$ \delta I (\cS : \cF) = I(\cS:\cF) - I(\{ \ket 0, \ket 1 \}:\cF) = 0 \ , \eqno(4.29) $$
and one can safely ascribe probabilities to correlations between $\cS$ and $\cF$ in the pointer basis
of $\cS$ singled out by the {\tt c-not} ``dynamics''. Discord would appear only if all of $\cE$ got
included, as then $I(\cS:\cF)=2H(\cS)$, twice the classical information marked by the plateau of the PIP.

By the same token, as soon as decoherence sets in, $H(\cS)=H(\cS,\cF)$ for any fragment $\cF$ that 
leaves enough of the rest of the environment $ \cE/ \cF$ to einselect pointer states in $\cS$. Consequently,
$$ I(\cS : \cF) =  I(\{ \ket 0, \ket 1 \}:\cF)  = H(\cF) \ , \eqno(4.30) $$
as well as;
$$ H(\cF) = H(\cS d \cF) \ , \eqno(4.31)$$
providing simple illustration of Eqs. (4.19)-(4.22).

\subsubsection{A spin system in a spin environment}

A straightforward generalization of controlled not gates and qubits is a model with a central spin system 
interacting with the environment of many other spins. Several different versions of dynamics of models
with this general structure were studied as examples of quantum Darwinism (Ollivier, Poulin and Zurek, 2004, 2005; Blume-Kohout and Zurek, 2005, 2006). 

Basic conclusions confirm that decoherence can indeed imprint multiple copies of the preferred 
observable onto the environment. Given the example of {\tt c-not}'s and qubits, this is no surprise. 
Copies of pointer states of $\cS$ are, of course, no longer perfect: A single subsystem of the 
environment is no longer perfectly correlated with (and, hence, does not completely decohere) 
the system. It is therefore also usually impossible for a single subsystem of $\cE$ to supply all the information about $\cS$. Nevertheless, when the environment is sufficiently large, asymptotic form of 
$I(\cS : \cF)$ has --
as a function of the size of the fragment $\cF$ -- the same character we have already encountered with {\tt c-not}'s: A steep rise (where in accord with Eq. (4.19), every bit of information 
stored in $\cF$ can be regarded as the mutual information ``about $\cS$'') followed by a plateau
(where the information only confirms what is already known). This is clearly seen in Fig. 4:
Only when the environment is too small to convincingly decohere the system, PIP
does not have a plateau.

{\begin{figure}[tb]
\begin{tabular}{l}
\vspace{-0.15in} 
\includegraphics[width=\FCW]{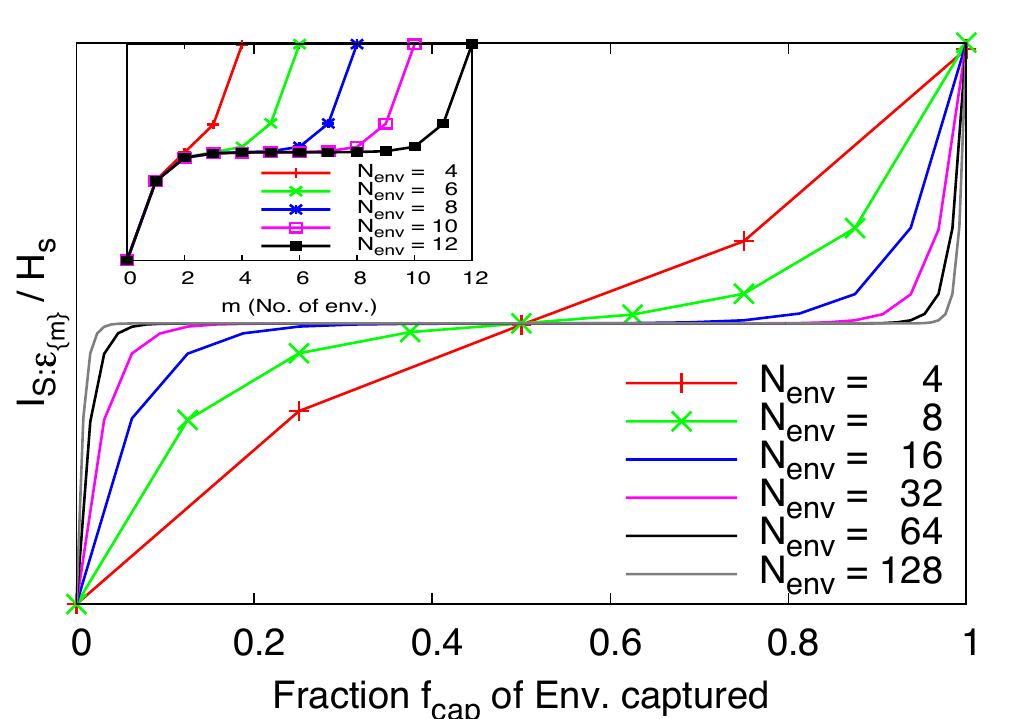}\\
\end{tabular}
\caption{Scaled partial information plot for a qutrit system coupled to $N=4\ldots128$ qutrit 
environments plotted against the size of the fraction $f$ (see Blume-Kohout and Zurek, 2006, 
for details). As the number of the environment subsystems increases, redundancy grows, which 
is reflected by increasing sharpness of the initial part of the plot. In this initial part of the plot 
appropriate re-scaling and assumptions of typicality bring out universal behavior that reflects 
Eq. (4.19). This is illustrated in the inset (Blume-Kohout and Zurek, 2005), which depicts rescaled 
entropy of a qubit plotted against the number of the environment qubits (rather than their fraction). Clearly, intercepting few 
spins in the environment (or few photons) provides all the essential new information that most 
of the other spins (or photons) only confirm. Elongation of the plateau is a symptom of the increase 
of redundancy. $R_\delta$ is the length of the PIP measured in units defined by the size 
(say, in the number of environment subsystems needed $m_\delta$, see Fig. 3) of the part  
of the PIP that corresponds to $I(\cS:\cF)$ rising from 0 to $(1-\delta) H(\cS)$. 
}
\label{SpinPIP}
\end{figure}

As we have seen with {\tt c-not}'s and qubits, and as will be often the case with photon environment, 
the system decoheres as soon as a single copy of its state is imprinted with a reasonable accuracy 
in $\cE$, and few such imprints will establish the initial rising part of the PIP. However, when new 
subsystems of $\cE$ become correlated with $\cS$, the size of the plateau increases and its elongation 
occurs without any real change to the early part of the PIP. Thus, the number of copies of the information 
$\cE$ has about $\cS$ can grow long after the system was decohered.

What is this information about? In systems such as spins with discrete observables one can prove rigorously that it is about pointer states. The proof was first given in the idealized case of perfect environmental record, and it was extended to the case of imperfect records 
a year later (Ollivier, Poulin, and Zurek, 2004; 2005). 

The key question is: What are the states $\{\ket {\sigma_k}\}$ of the system (or what is the set
of the corresponding observables $\{ {\bf O} \}$) that is completely ($I(\{\ket {\sigma_k}\}  : \cF) \approx H(\cS)$)
and redundantly ($R_{\delta}(\{\ket {\sigma_k}\}) \gg 1$ imprinted on the environment. Rigorous 
statement of the relevant theorem confirms Bohr's suspicion that rigor and clarity are complementary:
We quote (Ollivier, Poulin, and Zurek, 2004): {\it ``The set $\{ {\bf  O} \}$ is characterized by the unique 
observable ${\bf \Pi}$ called by definition `the maximally refined observable'. The information about any 
other observable $\sigma$ in $\{ {\bf  O} \}$ is obtainable from a fragment $\cF$ of $\cE$ is equivalent to the 
information that can be obtained through its correlations with the maximally refined ${\bf \Pi}$.''}

In other words, the most efficient way to predict outcome of measurement of {\it any} observable of the
system indirectly -- from the imprint in a part $\cF$ of the environment -- is to find out from $\cF$ about 
the pointer observable ${\bf \Pi}$ of the system, and then hope that the observable of interest $\sigma$ is
correlated with ${\bf \Pi}$. It goes without saying that no information can be
obtained in this way about observables complementary to ${\bf \Pi}$.

\begin{figure*}[tb] 

\center \includegraphics[width=5.8in]{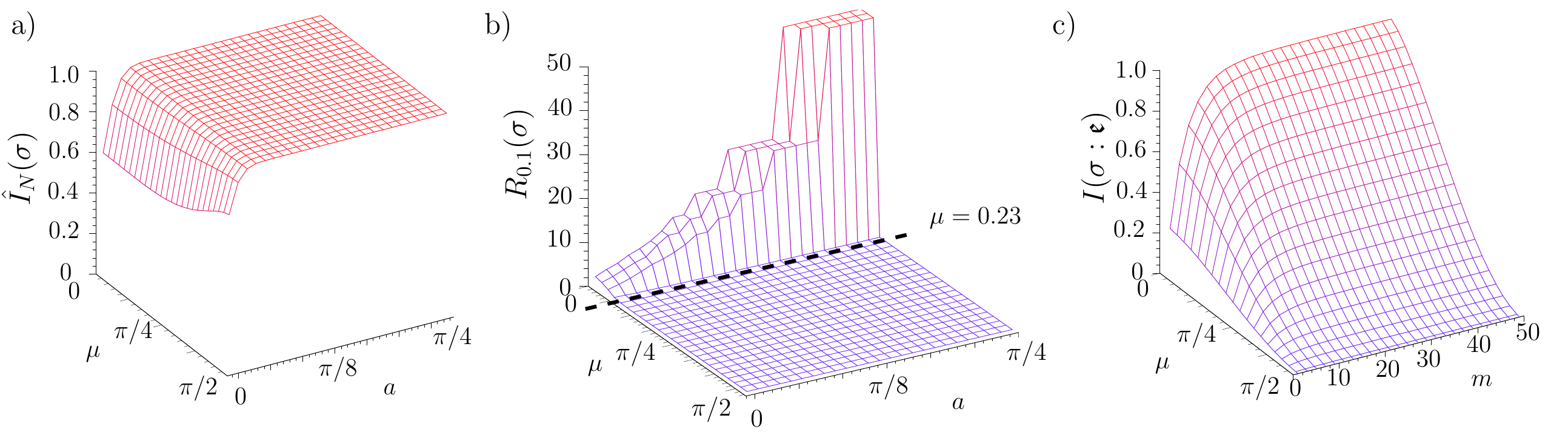}

\caption{Quantum Darwinism illustrated using a simple model of decoherence
 (Ollivier, Poulin, and Zurek, 2004). The system $\cS$, a spin-$\frac 12$ particle,
  interacts with $N=50$ two-dimensional subsystems of the environment
  through ${\bf H_{\cS\cE}} = \sum_{k=1}^N g_k \sigma_z^{\cS}
  \otimes\sigma_y^{\cE_k}$ for a time $t$. The initial state of
  $\mathcal S\otimes\cE$ is $\frac{1}{\sqrt 2}(\ket 0 + \ket 1)
  \otimes \ket 0 ^{\cE_1}\otimes\ldots\otimes\ket 0^{\cE_N}$. 
  Couplings $g_k$ are selected randomly with uniform distribution
  in the interval (0,1]. All the
  plotted quantities are function of the system's observable
  $\sigma(\mu) = \cos(\mu)\sigma_z + \sin(\mu)\sigma_x$, where $\mu$
  is the angle between its eigenstates and the pointer states of
  $\cS$---here the eigenstates of $\sigma_z^\cS$. {\bf a)}~Information
  acquired by the optimal measurement on the whole environment, 
  $\hat I_N(\sigma)$, as a function of the inferred observable $\sigma(\mu)$ 
  and the average interaction action $\langle g_k t \rangle = a$.
   A large amount of information is accessible in the {\em whole}
  environment for any observables $\sigma(\mu)$ except when the
  action $a$ is very small. Thus, complete imprinting of
  an observable of $\cS$ in $\cE$ is not sufficient to claim
  objectivity. {\bf b)}~Redundancy of the information about the system
  as a function of the inferred observable $\sigma(\mu)$ and the
  average action $\langle g_k t \rangle= a$. It is measured by
  $R_{\delta=0.1}(\sigma)$, which counts the number of times 90\% of
  the total information can be ``read off'' independently by measuring
  distinct fragments of the environment. For all values of the action
  $\langle g_k t \rangle =a$, redundant imprinting is sharply peaked around the
  pointer observable. Redundancy is a very selective criterion. The
  number of copies of relevant information is high only for the
  observables $\sigma(\mu)$ falling inside the theoretical bound (see
  text) indicated by the dashed line.  {\bf c)}~Information about
  $\sigma(\mu)$ extracted by an observer restricted to local random
  measurements on $m$ environmental subsystems.  The interaction
  action $a_k = g_k t $ is randomly chosen in $[0,\pi/4]$ for each
  $k$. Because of redundancy, pointer states---and only pointer
  states---can be found out through this far-from-optimal measurement
  strategy. Information about any other observable $\sigma(\mu)$ is
  restricted by our theorem to be equal to the information brought
  about it by the pointer observable $\sigma_z^\cS$.}
\label{opz}
\end{figure*}

We can illustrate this preeminence of the pointer observable on a simple model: a single spin $\frac 1 2$
interacting with a collection of $N$ such spins. As seen in Fig. 5a {\it environment as a whole
contains information about any observable of $\cS$}. Preferred role of the pointer observable becomes
apparent only when one seeks observables that are recorded {\it redundantly} in $\cE$. Figure 5b
shows that only the pointer observable ${\bf \Pi}= \sigma_z$ (and observables that are nearly parallel
to it) are recorded redundantly. The ``ridge of redundancy'' is strikingly sharp!

Comparison of Figure 5a and 5b also shows that redundancy of $ \sigma_z$ increases long 
after environment as a whole is strongly entangled with $\cS$. This is seen in a steady rise of $R_\delta$
with the action. 

Thus, in a more realistic model that {\tt cnot}'s and qubits we have seen characteristics (i) - (iii) of quantum Darwinism: (i) decoherence begets redundancy which can (ii) continue
to increase after decoherence has already happened. Moreover, (iii) both decoherence and quantum
Darwinism single out the same pointer observable. 

\subsubsection{Quantum Darwinism in Quantum Brownian Motion (QBM)}

Evolution of a single harmonic oscillator (the system) coupled through its coordinate with a collection 
of many harmonic oscillators (the environment) is a well known exactly solvable model 
(Feynmann and Vernon, 1963; Dekker, 1977; Caldeira and Leggett, 1983; Unruh and Zurek, 1989; Hu, Paz, and Zhang, 1992). 
In contrast to spin models (where exact and orthogonal pointer states can be often identified) 
preferred states selected by their predictability are Gaussian minimum uncertainty
wavepackets (Zurek, Habib, and Paz, 1993; Tegmark and Shapiro, 1994; Gallis, 1996). So, while decoherence in this
model is well understood, quantum Darwinism -- where the focus is not on
$\cS$, but on its relation to a fragment $\cF$ of $\cE$ -- presents one with novel challenges.

Here we summarize results obtained recently (Blume-Kohout and Zurek, 2007)
under the assumption that fragments of the environment 
are ``typical'' subsets of its oscillators -- that is, subsets of oscillators with the same spectral density 
as the whole $\cE$. The QBM Hamiltonian:
$$\Ham = \Hsys + \frac12
      \sum_{\omega}{\left(\frac{\penv{\omega}^2}{\Menv{\omega}}
     + \Menv{\omega}\omega^2 \xenv{\omega}^2\right)}
+ \xsys\sum_{\omega}{\CC{\omega}\xenv{\omega}} \eqno(4.32)$$
describes a collection of the environment oscillators coupled to the harmonic oscillator system with:
$$\Hsys = (\frac{\psys^2}{\Msys} + \Msys\omegabare^2 \xsys^2)/2 \eqno(4.33)$$
the environmental coordinates $\xenv{\omega}$ and $\penv{\omega}$ describe a single band (oscillator) $\Env_\omega$.  
As usual, the bath is defined by its spectral density,
$I(\omega) = \sum_{n}{\delta\left(\omega-\omegaenv{n}\right) 
            \frac{\CC{n}^2}{2\Menv{n}\omegaenv{n}}}, \label{eqSpectralDensity}$
which quantifies the coupling between $\Sys$ and each band of $\Env$.  We consider an
\emph{ohmic} bath with a cutoff $\Lambda$:
$I(\omega) = \frac{2\Msys\gamma_0}{\pi}\omega$
for $\omega\in[0\ldots\Lambda]$.
We adopt a sharp cutoff (rather than the usual smooth rolloff) to simplify numerics.
Each coupling is a differential element,
$\diff\CC{\omega}^2 = \frac{4\Msys\Menv{\omega}\gamma_0}{\pi}\omega^2\diff\omega$
for $\omega\in[0\ldots\Lambda]$.
For numerics, we divide $[0\ldots\Lambda]$ into discrete
bands of width $\Delta\omega$, which approximates the exact model
well up to a time $\tau_{rec} \sim
\frac{2\pi}{\Delta\omega}$.

We initialize the system in a squeezed coherent state, and $\Env$ in its ground state.
QBM's linear dynamics preserve the Gaussian nature of the state, which
can be described by its mean and variance:
$$
\vec{z} = \left(\begin{array}{c}\expect{x}\\ \expect{p}\end{array}\right)\mbox{\ ;\ }
\Delta = \left(\begin{array}{cc}\Delta x^2 & \Delta xp \\
 \Delta xp & \Delta p^2\end{array}\right). 
 $$
Its entropy is a function of its \emph{squared symplectic area},
$$ a^2 = \left(\frac{\hbar}{2}\right)^{-2}\det(\Delta) \label{eqSquaredAreaDef} \ ;  \eqno(4.34)$$
$$
\Hh(a) = \frac12\left(\begin{array}{l}\ (a+1)\ln(a+1)\\-(a-1)\ln(a-1)\end{array}\right)-\ln2
\approx \ln\left(\frac{e}{2}a\right), \eqno(4.35)$$
where $e$ is Euler's constant, and the approximation 
is excellent for $a>2$.  For multi-mode states, numerics yield $\Hh(\rho)$ exactly as a sum
over $\Delta$'s symplectic eigenvalues (Serafini et al., 2004), but our theory approximates a collection of oscillators as a single mode with a single $a^2$.

\begin{figure}[tb]
\begin{tabular}{l}
\vspace{-0.15in} \includegraphics[width=\FCW]{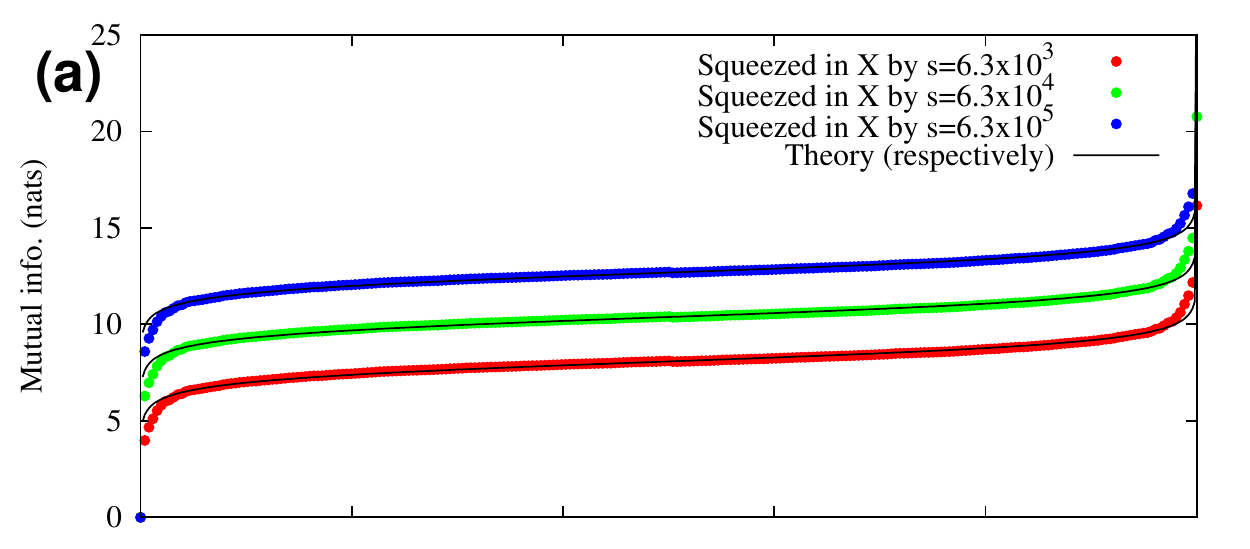}\\
\includegraphics[width=\FCW]{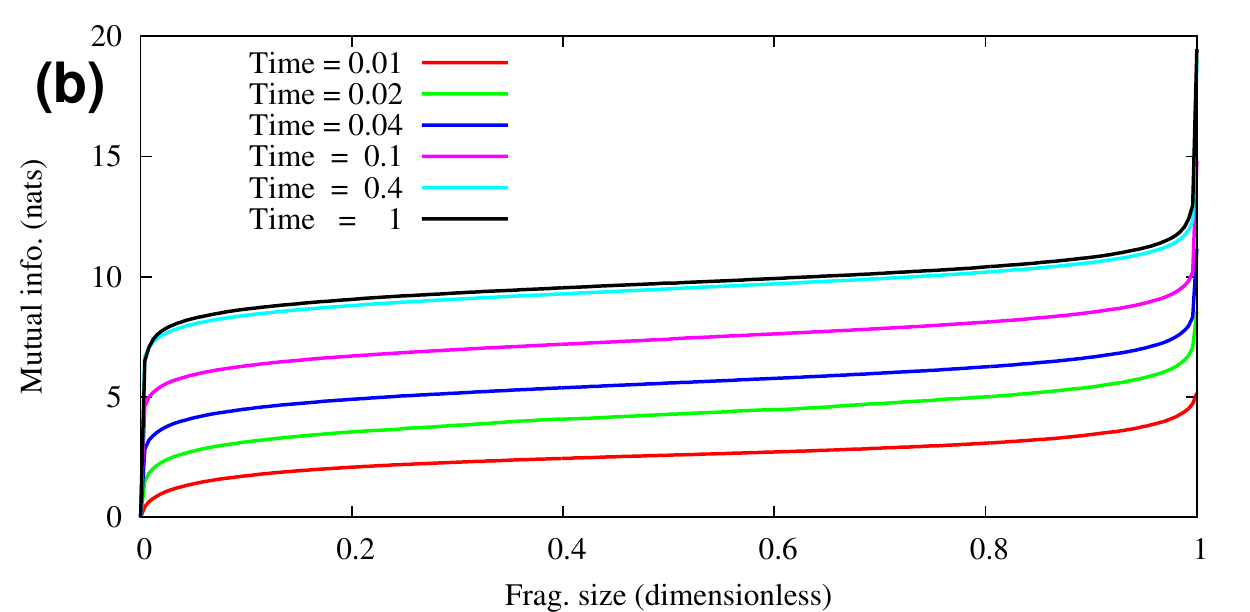}\\
\end{tabular}
\caption{\emph{Partial information plots for quantum Darwinism in QBM} (Blume-Kohout and Zurek, 2007). To obtain graphs above, the system $\Sys$ was initialized in an $x$-squeezed state, which decoheres as it evolves into a superposition of localized states.  Plot \textbf{(a)} shows PIPs for three fully-decohered ($t=4$) states with different squeezing.  Small fragments of $\Env$ provides most of the available information about $\Sys$; squeezing changes the amount of \emph{redundant} information without changing the PIP's shape.  The numerics agree with our simple theory.  Plot \textbf{(b)} tracks one state as decoherence progresses.  PIPs' shape is invariant; time only changes the redundancy of information.}
\label{figPIPs}
\end{figure}

PIPs (Fig. 6) show information about $\Sys$ stored in $\Env$.
$I(\cS :\cF )$ rises rapidly as the fragment's size ($f$) increases from zero,
then flattens for larger fragments.  
Most -- all but $\sim1$ nat -- of $\Hs$ is recorded redundantly in $\cE$.
When $\Sys$ is macroscopic, this \emph{non-redundant information}
is dwarfed by the total amount of information lost to $\Env$.

Calculations simplify in the macroscopic limit where the mass of the system is large compared to 
masses of the environment oscillators. This regime (of obvious interest to the quantum - classical 
transition) allows for analytic treatment based on the Born-Oppenheimer approximation: 
Massive system follows its classical trajectory, largely unaffected by $\cE$. The environment will, 
however, decohere a system that starts in a superposition of such trajectories. In the process, $\cE$ 
that starts in the vaccum will become imprinted with the information about position of $\cS$. This leads 
to consequences for partial information plots and for redundancy we shall now discuss.

The basic observation is that the area of the $1-\sigma$ contours 
in phase space determines entropy.  As a result of decoherence, the area corresponding to the state of the system will increase by $\delta a^2_\Sys$. This is caused by the entanglement with the environment,
so the entropies and areas of environment fragments increase as well.  When $\Frag$ contains a 
randomly selected fraction $f$ of $\Env$, $\rho_{\Frag}$'s squared area is 
$a^2_{\Frag} = 1 + f\delta a^2_\Sys$, and that of $\rho_{\Sys\Frag}$ is
$a^2_{\Sys\Frag} = 1 + (1-f)\delta a^2_\Sys$.  Applying Eq. (4.35) (where $\delta a^2_\Sys \gg 1$) yields:
$$I(\cS : \cF) \approx \Hh(\Sys) + \frac12\ln\left(\frac{f}{1-f}\right) \eqno(4.36)$$
This ``universal'' $I(\cS:\cF)$ is valid for significantly delocalized initial states of $\cS$. It is a good approximation everywhere except very near $f=0$ and $f=1$ (where it would predict singular behavior).
It has  a classical plateau at $H_{\cS}$ which rises as decoherence increases entropy of the system.
In contrast to PIP's we have seen before, adding more oscillators to the environment does not simply 
extend the plateau: The shape of $I(\cS:\cF)$ is only a function of $f$ and so it is invariant under 
enlargement of $\cE$. This invariance is caused by the fact that adding more oscillators to 
the environment increases entropy of the system (while in case of spins $H(\cS)$ was limited 
by Shannon entropy of the pointer observable). 

\begin{figure*}[tb!]
\setlength{\unitlength}{\textwidth}
\begin{picture}(1,0.4)
\put(0,0){\includegraphics[width=0.6\textwidth]{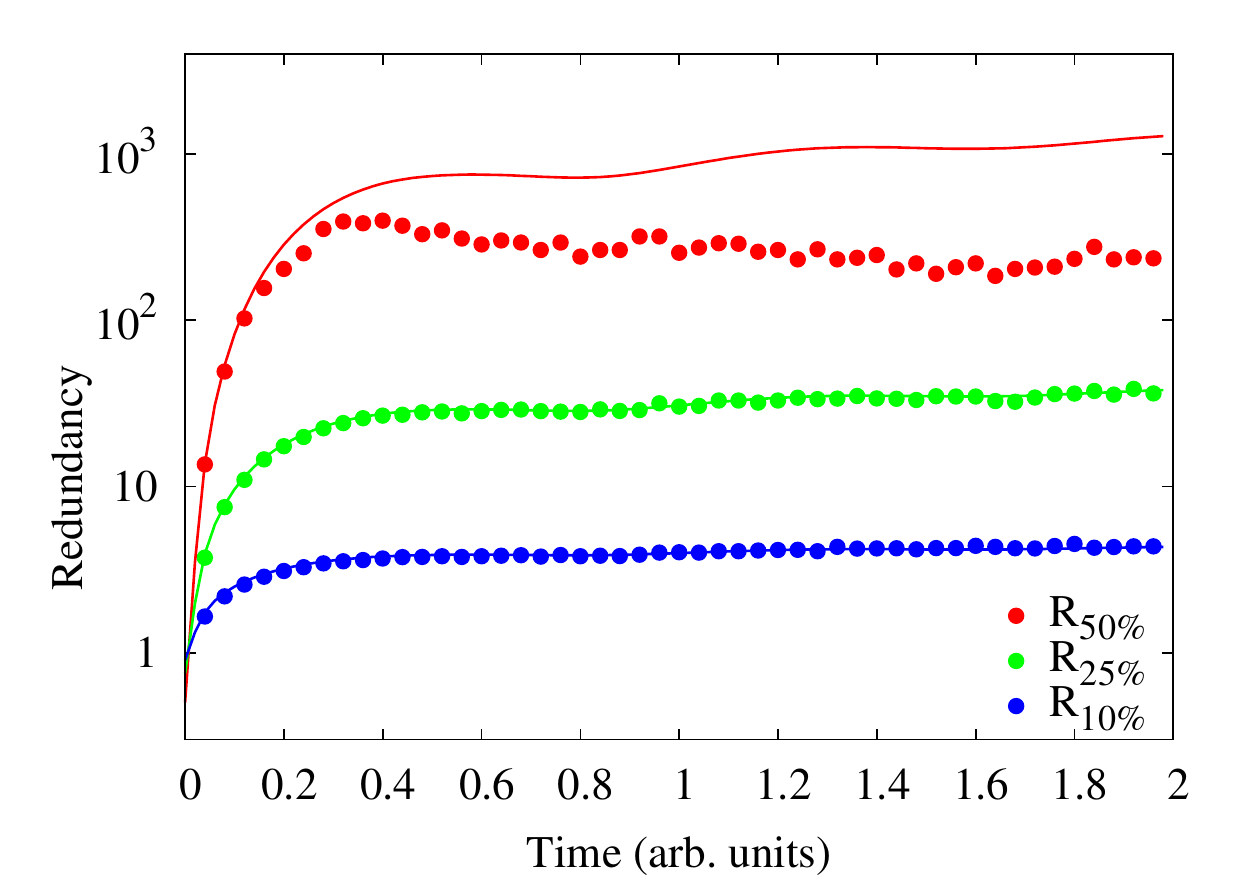}} \put(0.52,0.37){\mbox{\large\textbf{(a)}}}
\put(0.575,0.275){\includegraphics[width=0.42\textwidth]{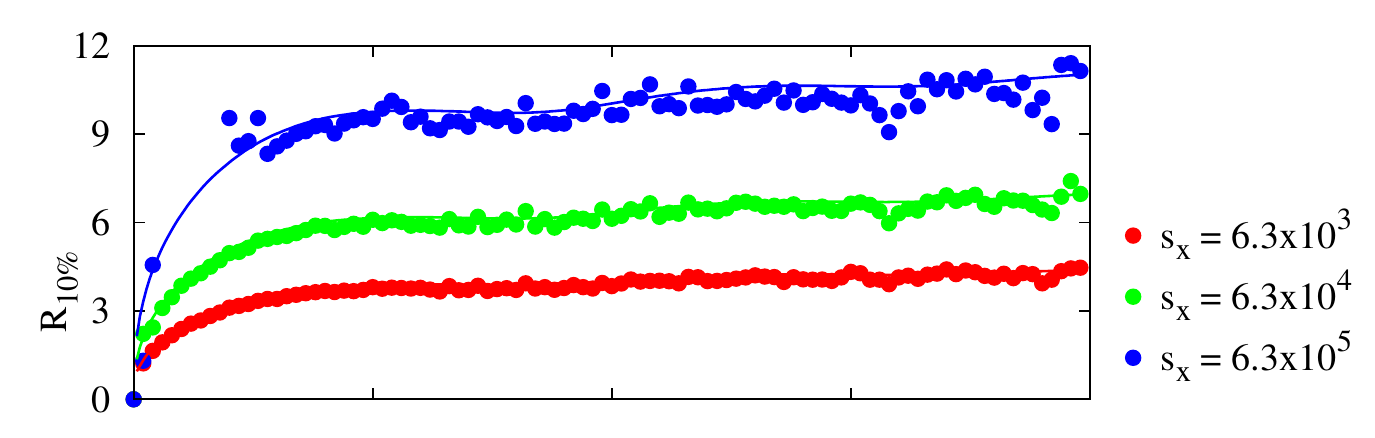}} \put(0.91,0.375){\mbox{\large\textbf{(b)}}}
\put(0.575,0.14){\includegraphics[width=0.42\textwidth]{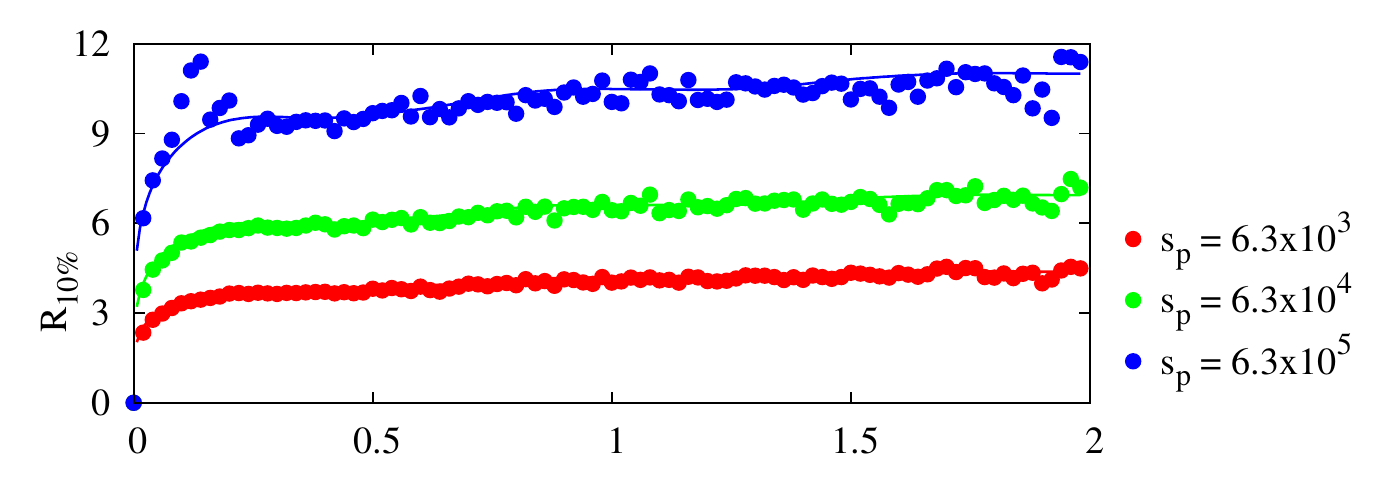}} \put(0.91,0.253){\mbox{\large\textbf{(c)}}}
\put(0.575,0){\includegraphics[width=0.42\textwidth]{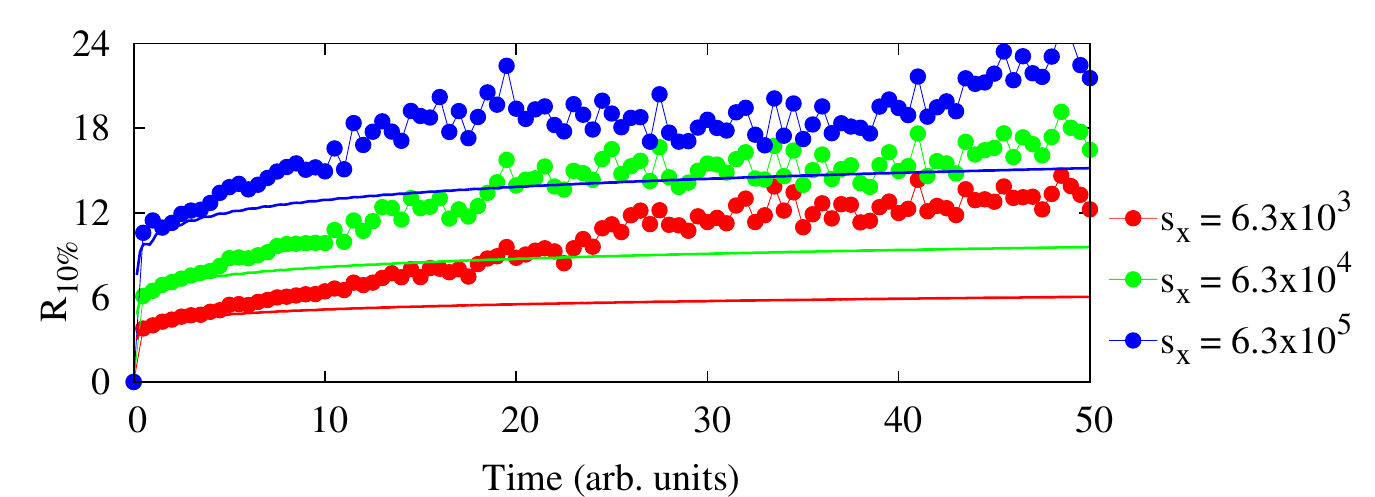}} \put(0.91,0.12){\mbox{\large\textbf{(d)}}}
\end{picture}
\caption{\emph{Delocalized states of a decohering oscillator ($\Sys$) are redundantly recorded by the environment ($\Env$).}  Plot \textbf{(a)} shows redundancy ($R_\delta$) vs. imprecision ($\delta$), when $\ket{\psi(0)}$ is squeezed in $x$ by $s_x = 6.3\times10^3$.  Plots \textbf{(b-d)} show $R_{10\%}$ -- redundancy of 90\% of the available information -- vs. initial squeezing ($s_x$ or $s_p$).  Dots denote numerics; lines -- theory. $\Sys$ has mass $\Msys=1000$, $\omegasys=4$.  $\Env$ comprises oscillators with $\omega\in[0\ldots16]$ and mass $\Menv{}=1$.  The frictional (coupling) frequency is $\gamma=\frac{1}{40}$. Redundancy develops with decoherence: $p$-squeezed states [plot \textbf{(c)}] decohere almost instantly, while $x$-squeezed states [plot \textbf{(b)}] decohere as a $\frac\pi2$ rotation transforms them into $p$-squeezed states.  Redundancy persists thereafter [plot \textbf{(d)}]; dissipation intrudes by $t\sim O(\gamma^{-1})$, causing $R_{10\%}$ to rise above simple theory.  
Redundancy increases \emph{exponentially} -- as $R_\delta \approx s^{2\delta}$ -- with imprecision [plot \textbf{(a)}]. So, while $R_{\delta} \sim 10$ may seem modest, $\delta=0.1$ implies \emph{very} precise knowledge (resolution around 3 ground-state widths) of $\Sys$. This is half an order of magnitude better than a recent record (Lahaye et al., 2004) for measuring a micromechanical oscillator.  At $\delta \sim 0.5$ -- resolving $\sim \sqrt{s}$ different locations within the wavepacket -- $R_{50\%}\gtrsim 10^3$ (our maximum numerical resolution; see Blume-Kohout and Zurek, 2007).
}
\label{figRedundancy}
\end{figure*}
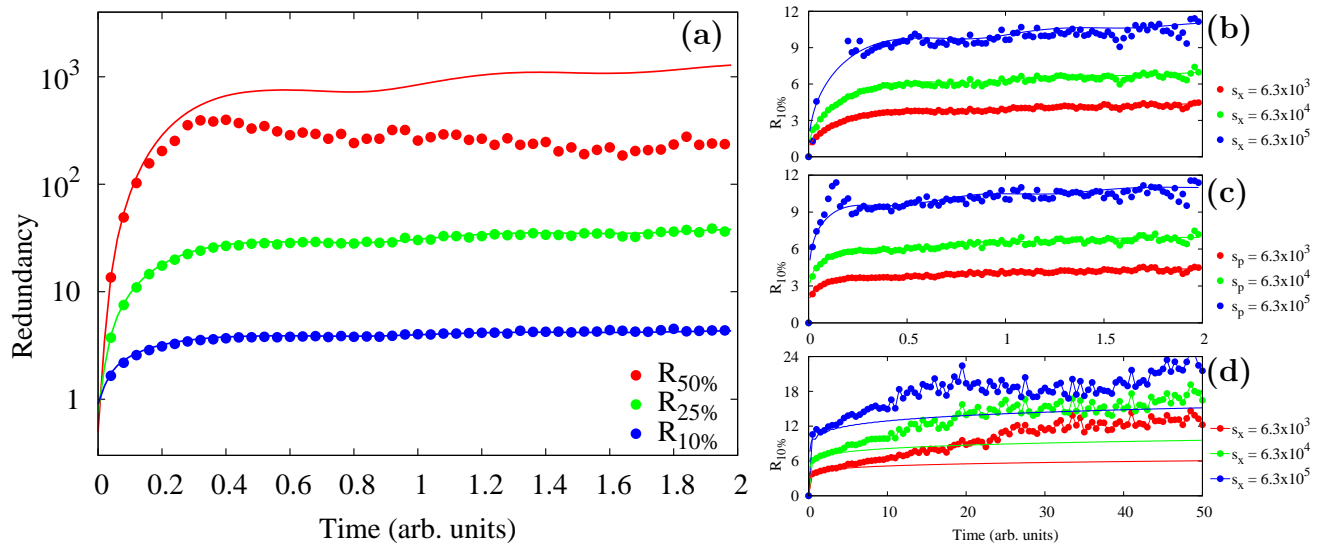

When the above equation for $I(\cS:\cF)$ is solved for $f_\delta$ one arrives at the estimate:
$$ R_\delta \approx e^{2\delta\Hs} \approx s^{2\delta}. \eqno(4.37)$$
The last equality above follows because an $s$-squeezed state decoheres to a mixed state with 
$H_\Sys \approx \ln s$. This simple last equation for $R_{\delta}$ holds where it matters -- after
decoherence but before relaxation begins to force the system to spiral down towards its ground state. 
As  trajectories decay, plateau flattens compared to what Eq. (4.36) would predict. This will initially 
increase redundancy $R_{\delta}$ above the values attained after decoherence (see Fig. 7c).
Eventually, as the whole $\cS\cE$ equilibrates, the system will spiral down to occupy mixture 
of low-lying number eigenstates, and $R_\delta$ will decrease. 

In this perhaps most realistic (but still idealized) QBM model we confirm that 
decoherence dynamics leads to quantum Darwinism. While buildup of redundancy takes longer than the initial destruction of quantum coherence, various time-dependent processes (such as the increase of
redundancy caused by dissipation) are still to be investigated in detail. Moreover, localized states
favored by einselection are redundantly recorded by $\cE$. So, quantum Darwinism in QBM confirms 
features of decoherence we have anticipated earlier. On the other hand, we have found an interesting tradeoff  between redundancy and imprecisoon $\delta$, Eq. (4.37). Objectivity (as measured by redundancy)
comes at the price of accuracy.

We also note that the limit of effective decoherence we have described before holds
in case of QBM. This follows, in effect, from the fact that $a^2_{\Sys\Frag} = 1 + (1-f)\delta a^2_\Sys$
approaches $a^2_{\Sys} = 1 + \delta a^2_\Sys$ for small $f$. Consequently, it is evident that 
$H(\cS)-H(\cS,\cE/\cF) \approx 0$, and the mutual information $I(\cS:\cF)$ is given by $H(\cF)$.
This is not obvious from the simple scale invariant equation (4.37) above. 

\subsection{Summary: Environment as a Witness}

States in classical physics were ``real'': Their {\it objective existence} was established operationally -- 
they could be found out by an initially ignorant observer without getting perturbed in the measurement 
process. Hence, they existed independently of what was known about them. 

Information was, by 
contrast, ``not real''. This was suggested by immunity of classical states to measurements. Information 
was what observer knew subjectively, a mere shadow of the real state, irrelevant for physics. 

This dismissive view of information run into problems when classical Universe of Newton confronted thermodynamics. Clash of these two paradigms led to Maxwell's demon, and is implicated in the origins 
of the arrow of time. The specter of information was and still is haunting physics. The seemingly 
unphysical record state was beginning to play a role reserved for the real state!

We have just seen how, in quantum setting, information and existence become interdependent.
The real state is defined and made objective by what is known about it -- by the information. 
``It from bit'' comes to mind (Wheeler, 1990). The main new ingredient is the environment. 
$\cE$ acts as a witness of the quantum state of the ``object of interest''. It has information -- many 
copies of information -- about $\cS$. So, to quote another famous dictum, ``information is physical''
(Landauer, 1991). It must reside somewhere (e.g. in the environment). And the existence of evidence 
has its legal consequences. 

The role of $\cE$ in quantum Darwinism is not 
that of an innocent bystander, who simply reports what has happened. Rather, the environment is an 
accomplice in the ``crime'', selecting and transforming some of the fragile ``epistemic'' quantum states 
into robust, objectively existing classical states: Environment - induced decoherence invalidates quantum principle of superposition, leading to einselection (which censors Hilbert space). 
Information transfer associated with it selects preferred pointer states, and banishes their superpositions.

Moreover, testimony $\cE$ gives is biased -- it can reveal only the very same pointer states $\cE$ has
helped select. Operational criterion for objectivity is the ability to find out a state without disturbing it. 
According to this operational definition of objective existence, pointer states exist in more or less 
the same way their classical counterparts did: They can be found out without getting perturbed by anyone who examines one of the multiple copies of the record of $\cS$ ``on display'' in the environment. 

\section{Discussion: Existential Interpretation}

There are two key ideas in Everett's writings. The first one is to let quantum theory dictate its 
own interpretation. We took this ``let quantum be quantum'' point very seriously. The second message
(that often dominates in popular accounts) is the Many Worlds mythology. In contrast ``let quantum be quantum'' it is less clear what it means, so -- in the opinion of this author -- there is less reason 
to take it at face value. 

In search for relation between quantum formalism and the real world we have weaved together several 
ideas that are very quantum to arrive at the {\it existential interpretation}. Its essence is the operational definition of objective existence of physical states: To exist, a state must, at the very least, 
persist or evolve predictably in spite of the immersion of the system in its environment. Predictability 
is the key to einselection. Objective existence requires more: It should be possible 
to find out a state without perturbing it -- without threatening its existence.

Let us briefly recapitulate how objective existence arises in the quantum setting: We started with 
axioms (i) and (ii) that sum up mathematics of quantum theory: They impose quantum principle of 
superposition, and demand unitarity, but make no connection with the ``real world''. Addition of predictability postulate (axiom (iii), the only uncontroversial measurement axiom), and recognition that our Universe consists of systems (axiom (o)) immediately leads to {\it preferred sets of pointer states}. 
In the context of Everett's relative state interpretation (that explains why a quantum observer perceives a 
definite outcome) this conclusions settles the issue of ``collapse''. It justifies Hermitean nature of
quantum observables and explains breaking of unitary symmetry, the crux of the collapse axiom (iv).

Our next task was to understand {\it the origin of probabilities and Born's rule}, axiom (v). We 
have done this without appealing to decoherence (as this would have meant circularity in the 
derivation). Nevertheless, decoherence -- inspired view is reflected in the envariant 
approach: To attribute probabilities to pointer states we first showed how to ``get decoherence without using tools of decoherence'' -- to demonstrate that relative phases between outcomes
do not matter. Envariance provides the answer, but -- strictly speaking -- only for Schmidt states. 

We take this to mean that usual rules of the probability calculus will strictly hold for pointer states only 
after they have decohered. Pointer states -- Schmidt states coincidence is expected to be very good 
indeed: Probabilities one has in mind ultimately refer to pointer states of measuring or recording devices. These are usually macroscopic, so their interaction with the environment will quickly align 
Schmidt basis with pointer states. Born's rule (with all the consequences for the frequencies of events) immediately follows. 

Probabilities derived from envariance are objective: They reflect experimentally testable 
symmetry of the global state (usually involving the measured system $\cS$, the apparatus $\cA$, 
and its environment $\cE$). Before observer interacts with $\cA$ he will be ignorant of the outcome, 
but will know the set of pointer states -- the menu of possibilities. This ignorance reflects objective symmetries of the global state of $\cS\cA\cE$. These symmteries of the global state lead to Born's rule. 

Last question left to address was the origin of {\it objective existence} in the quantum world. We started 
by noting that in contrast to fragile arbitrary superpositions in the Hilbert space of the system, pointer 
states are robust. They are associated with operators (e.g., a completely positive map that embodies decoherence process). Therefore, pointer states can be ``found out'': In contrast to states (which are 
fragile), operators in quantum theory are robust, and an unknown quantum operator can be usually
determined. In the simplest case pointer states are the eigenstates 
of the interaction Hamiltonian (Zurek, 1981; 1991). As noted by Aharonov, Anandan, and Vaidman (1993; see also Unruh, 1994), this changes their ontological status. But this account loosely based on ``protective measurements'' (while not incorrect in principle) does not capture what happens in practice. 

In the real world observers find out pointer states of systems not by carefully investigating interaction 
Hamiltonians through protective measurements (as if they were dealing with fragile endangered 
species), but by letting natural selection take its course: Pointer states are the ``robust species'', adapted to their environments. They survive intact its monitoring. More importantly, multiple records 
about $\cS$ are deposited in $\cE$. They favor pointer states, which are the ``fittest'' in 
the Darwinian sense -- they can survive and multiply. 

There is an extent to which ``it had to be so'': In order to make one more copy 
of the original one needs to preserve the original. But there is a more subtle part of this relation between 
decoherence, einselection, and quantum Darwinism. Hamiltonians of interaction that allow for copying 
of certain observables necessarily leave them unperturbed. This conspiracy was noticed early: It is the
basis of the commutation criterion for pointer observables (Zurek, 1981). In effect, $\bf {H}_{\cS\cE}$ 
is a function of some observable ${\bf \Lambda}$ of the system, it will also necessarily commute with it, $[\bf { H}_{\cS\cE}, ~{\bf \Lambda}]=0$. 


Existential interpretation of quantum theory assigns ``relatively objective existence'' (Zurek, 1998b)
-- key to effective classicality -- to widely broadcast quantum states. It is obviously consistent with the
relative state interpretation: Redundancy of records disseminated throughout the environment 
supplies a natural definition of branches that are classical in the sense that an observer can find out macroscopic features of his branch and stay on it,  rather than ``cut off the branch he is sitting on'' 
with his measurement. This is more than einselection, and much more than decoherence, although 
the key ingredient -- environment -- is still the key, and the key criterion is ``survival of the fittest'' -- 
immunity of pointer states to monitoring by $\cE$ reflected also in the predictability sieve. 
The role of $\cE$ is however upgraded from a passive ``quantum information dump'' to that of a 
communication channel. Information deposited in $\cE$ in the process of decoherence is not lost. Rather, it is stored there, in multiple copies, for all to see. The emphasis on information theoretic 
significance of quantum states cuts both ways: Environment - as - a - witness paradigm supplies 
operational definition of objective existence, and shows why and how pointer states can be found out 
without getting disrupted. However, it also shows that objective existence is not an intrinsic property of a quantum state, but that it arises along with -- and as a result of -- information transferred from
$\cS$ to $\cE$.

\subsection{Bohr, Everett, and Wheeler}

Careful reader will note that this paper has largely avoided issues of interpretation, focusing 
instead on consequences of quantum theory that are ``interpretation independent'', but may constrain 
interpretational options. This has certainly been our strategy: We have been led by quantum formalism 
to our conclusions, but these conclusions are largely beyond dispute. Our ``existential interpretation'' 
is in that sense not an interpretation -- it simply points out the consequences of quantum formalism and 
some additional rudimentary assumptions. 

It is nevertheless useful to see how the two principal interpretations of quantum theory -- Bohr's 
``Copenhagen Interpretation'' (CI) and Everett's ``Relative State Interpretation'' (RSI) fit within the 
constraints that we have derived above by acknowledging paramount role of the environment.
To anticipate the conclusion, we can do no better than to quote John Archibald Wheeler (1957), 
who, comparing CI with RSI, wrote: ``(1) The conceptual scheme of ``relative state'' quantum mechanics 
is completely different from the conceptual scheme of the conventional ``external observation'' form 
of quantum mechanics and (2) The conclusions from the new treatment correspond completely in 
familiar cases to the conclusions from the usual analysis.''

Bohr insisted on the classical part of the Universe to render outcomes of quantum measurements 
firm and objective. We have quantum Darwinism to accomplish that goal. Decoherence takes away quantumness of the system, but a system that is not quantum need not be immediately classical:
Objective nature of events arises only as a result of redundancy, and is reached only asymptotically,
in the limit of infinite redundancy, but large redundancy yields a very good approximation -- sort 
of like finite systems that have a critical point marking a phase transition which is, strictly speaking, 
precisely defined only in the infinite size limit. Indeed, Quantum Darwinism might be regarded 
as a purely quantum implementation of the ``irreversible act of amplification'' that was such an 
important element of CI. 

Physical significance of a quantum state in CI was purely epistemic (Bohr, 1928; Peres, 1993): 
States were only carriers of
information -- they correlated outcomes of measurements. Only the classical part of the Universe
existed in the sense we are used to. In the account we have given above there are really several 
different sorts of states. There are pure states -- vectors in the Hilbert space. But there are also objectively unknown states defined by the ``Facts''. They describe a subsystem and derive from 
pure state of the whole -- characterized by its symmetries. 
And there are states defined through the spectral decomposition of a quantum operator. Pointer states 
can be identified in this manner. 

Role of each of these states changes depending on the context -- provided by the information about it
as it is filtered by its environment. For instance, pure pointer state can be said to exists objectively 
when it has spawned -- through quantum Darwinism -- enough environmental progeny to be discovered 
without getting disturbed. But, in an isolated system, the very same state is clearly not objective for 
an observer who does not know it (and, hence, cannot measure it without re-preparing the system). 
On the other hand, when such an observer knows that this state is one of the eigenstates of an operator 
he can inspect at leisure, that pure state is much like an unknown classical state -- with
enough care and perseverance it can be ``found out''. 

Mixed states also have a meaning that depends on the context. Often observer knows enough to 
devise a measurement that will reveal to him one of the pointer states, but leave the mixed state 
of the system (although not necessarily the global state of the whole, e.g., of system and 
the environment that caused it to decohere) intact. Many observers can do this, or they can use 
environment as a witness. In either case, they will end up agreeing on the results. In this case mixed 
quantum state is in effect a probability distribution over effectively classical states.

This use of correlations as a source of consensus was recognized by Everett (1957a;b) and emphasized by Wheeler (1957). But, in contrast to CI that split Universe into only two domains -- quantum and classical -- we have seen that classicality is a matter of degree, and a matter of a criterion. 
For example, objectivity (which is in a sense the strongest criterion) is attained only asymptotically, in the 
limit of very large redundancy. It is clear why this is a good approximation in the case of macroscopic systems. But it is also clear that there are many intermediate stages on the way from quantum to classical, and that a system can be no longer quantum but be still far away from classical objective existence.

It is especially encouraging for the relative states point of view that the long - standing problem of 
the origin probabilities has an elegant solution that is very much ``relative state'' in spirit. We have 
relied on symmetries of entangled states. This allowed us to derive objective probabilities for individual 
events. We note that this is the first such {\it objective} derivation of probabilities not just in the quantum 
setting, but also in the history of the concept of probability. 

Envariant derivation of Born's rule is based on entanglement (which is at the heart of relative states 
approach). We have not followed either proposals that appeal directly to invariant measures on the Hilbert space (Everett, 1957a,~b), or attempts to derive Born's rule from frequencies by counting many worlds branches (Everett, 1957b; DeWitt, 1970, 1971; Graham 1973, Geroch, 1984): As noted by 
DeWitt (1971) and Kent (1990), Everett's appeal to invariance of Hilbert space measures makes 
no contact with physics, and makes even much less physical sense than the already very unphysical proof of Gleason (1957). And frequentist derivations are circular  -- in addition to counting branches they implicitly use Born's rule to dismiss ``maverick universes''. 

\subsection{Closing remarks}

Our strategy was to avoid purely interpretational issues and to focus instead on technical 
questions. They can be often answered in a definitive manner. In this way we have gained 
new insights into selection of preferred pointer states that go beyond decoherence, found out how 
probabilities arise from entanglement, and discovered that objectivity derives from redundancy. 

All of that fits well with the relative states point of view.
But there are also questions related the above technical developments but are at present 
less definite -- less technical -- in nature. We signal some of them here.

The first point concerns nature of quantum states, and its implications for interpretation.
One might regard states as purely epistemic (as did Bohr) or attribute to them ``existence''. 
Technical results described above suggest that truth lies somewhere between these two extremes. 
It is therefore not clear whether one is forced to attribute ``reality'' to all of the branches 
of the universal state vector. Indeed, such view combines a very quantum idea of a state 
in the Hilbert space with a very classical literal ontic interpretation of that concept. These two 
views of the state are incompatible. As we have emphasized, unknown quantum state cannot 
be found out. It can acquire objective existence only by ``advertising itself'' in the environment. 
This is obviously impossible for universal state vector -- the Universe has no environment. 

Objective existence can be acquired (via quantum Darwinism) only by a relatively small fraction of 
all degrees of freedom within the quantum Universe: The rest is needed to ``keep records''. Clearly, there is only a limited (if large) memory space available for this at any time. This limitation on the total 
memory available means that not all quantum states that exist or quantum events that happen now 
``really happens'' -- only a small fraction of what occurs will be still in the records in the future. 
So the finite memory capacity of the Universe implies indefiniteness of the present and impermanence 
of the past: To sum it up, one can extend John Wheeler's dictum  ``the past exists only insofar 
as it is recorded in the present'' and say ``whatever exists is there only insofar as it is recorded''.

\hocom{
In Section III we have, in effect,  derived reduced density matrices from envariance. 
Reduced density matrix is a complete description of a system. It does not betray origin 
of the ignorance it quantifies: An ``improper mixture'' caused by entanglement is locally 
indistinguishable from a ``proper mixture'' of well defined orthogonal pure states. 
Yet, some (see e.g. d'Espagnat, 1995) insist on the need to distinguish between 
``proper'' and ``improper'' mixtures. Indeed, the essence of proofs of  ``insolubility'' of 
the measurement problem is based on such a distinction. Quantum theory suggests 
precisely the opposite attitude: When it is impossible to tell proper from improper mixtures, 
there is no way for observers to know if they are discovering a predetermined outcome 
with a measurement, or ``collapsing'' some globally pure state. This quantum principle of 
equivalence applies also when there is an obvious preferred basis. In such a case 
(brought about e.g., by quantum Darwinism) observers are in effect dealing with 
a pre-collapsed state of the system -- the set of outcomes is fixed. Many observers can 
find out what is the specific outcome without invalidating each other's determinations. 
Quantum mechanics makes it impossible for them  to distinguish -- on the basis of 
the information accessible to them -- whether they have found out an effectively classical 
preexisting state, or a ``collapsed'' a global state. 
}

I would like to thank Robin Blume-Kohout, Fernando Cucchietti, Harold Ollivier, Juan Pablo Paz, 
and David Poulin for stimulating discussion.


\begin{thebibliography}{99}

\bibitem{1} Aharonov, Y, J. Anandan, and L. Vaidman, 1993, Phys. Rev. {\bf A47}, 4616.

\bibitem{Alb} Albrecht, A., 1992, Phys. Rev. {\bf D 46}, 550.

\bibitem{2} Auletta, G., 2000, {\it Foundations and Interpretation of Quantum Theory} (World Scientific,
Singapore).

\bibitem{3} Bacciagaluppi, G., 2004, in {\it The Stanford Encyclopedia 
of Philosophy}, E. N. Zalta, ed., on http://plato.stanford.edu/entries/qm-decoherence.

\bibitem{4} Baker, D. J., 2007,  Stud. Hist. Phil. Mod. Phys. {\bf 38}, 153.

\bibitem{5} Barnum, H., 2003, {\it No-signalling-based version of Zurek's derivation 
of quantum probabilities: A note on ``Environment-assisted invariance,
entanglement, and probabilities in quantum physics}, quant-ph/0312150

\bibitem{6} Barnum, H., C. M. Caves, J. Finkelstein, C. A. Fuchs, and 
R. Schack, 2000, {\it Proc. Roy. Soc.} London {\bf A456}, 1175.

\bibitem{7} Bennett, C. H., G. Brassard, and N. D. Mermin, 1992, Phys. Rev. Lett. {\bf 68}, 557.

\bibitem{8} Blume-Kohout, R., and Zurek, W. H., 2005, Found. Phys. {\bf 35}, 1857.

\bibitem{9} Blume-Kohout, R., and Zurek, W. H., 2006, Phys. Rev. {\bf A73}, 062310.

\bibitem{10} Blume-Kohout, R., and Zurek, W. H., 2007,  arXiv:0704.3615

\bibitem{11} Bohr, N., 1928, {\it Nature} {\bf 121}, 580; reprinted in Wheeler and Zurek, (1983).

\bibitem{12} Born, M., 1926, {\it Zeits. Phys.} {\bf 37}, 863; reprinted in Wheeler and Zurek, (1983).

\bibitem{13} Buniy, R. V., S. D. H. Hsu, and A. Zee, 2006, {\it Phys. Lett.} {\bf B640}, 219.

\bibitem{14} Caldeira, A. O., and A. J. Leggett, 1983, {\it Physica} A {\bf 121}, 587.

\bibitem{15} Caves, C. M., and Schack, R., 2005, {\it Ann. Phys.} {\bf 315}, 123.

\bibitem{16} Dekker, H., 1977, Phys. Rev. {\bf A16}, 2126.

\bibitem{17} Deutsch, D., 1985, Int. J. Theor. Phys. {\bf 24}, 1.

\bibitem{18} Deutsch, D., 1997, {\it The Fabric of Reality}, (Penguin, New York).

\bibitem{19} Deutsch, D., 1999, Proc. Roy. Soc. London, Series A {\bf 455} 3129. 

\bibitem{20} DeWitt,  B. S., 1970, Physics Today {\bf 23}, 30.

\bibitem{21} DeWitt,  B. S., 1971, in {\it Foundations of Quantum Mechanics}, B. d'Espagnat, ed.
(Academic Press, New York); reprinted on pp. 167-218 of DeWitt and Graham, (1973).

\bibitem{22} DeWitt, B. S., and N. Graham, eds., 1973, {\it The Many - Worlds Interpretation 
of Quantum Mechanics} (Princeton University Press, Princeton).

\bibitem{23} Dirac, P. A. M., 1958, {\it Quantum Mechanics} (Clarendon Press, Oxford).

\bibitem{24} Dieks, D., 1982, Phys. Lett. {\bf 92A}, 271.

\bibitem{25} Everett III, H., 1957a, Rev. Mod. Phys. {\bf 29}, 454; reprinted in Wheeler and Zurek, (1983).

\bibitem{26} Everett III, H., 1957b, Ph. D. Dissertation, Princeton University, reprinted in DeWitt 
and Graham (1973).

\bibitem{27} Fine, T. L., 1973, {\it Theories of Probability: 
An Examination of Foundations} (Academic Press, New York).

\bibitem{28} Feynman, R. P., and R. L. Vernon, 1963, {\it Ann. Phys.} (N. Y.) {\bf 192}, 368.

\bibitem{29} Forrester, A., 2007, {\it Stud. Hist. Phil. Mod. Phys.}, to appear; arXiv:quant-ph/0604133.

\bibitem{30} Gallis, M.~R., 1996, Phys.\ Rev. {\bf A53}, 655.

\bibitem{31} Geroch, R., 1984, No\^us {\bf 18} 617.

\bibitem{32} Gnedenko,  B. V., 1968, {\it The Theory of Probability} 
(Chelsea, New York).

\bibitem{33} Graham, N., 1973, {\it The Measurement of Relative Frequency} 
pp. 229-253 in B. S. DeWitt and N. Graham, in {\it The Many - Worlds Interpretation 
of Quantum Mechanics} (Princeton University Press).


\bibitem{H} Herbut, F., 2007, J. Phys. A. {\bf 40}, 5949.

\bibitem{34} Hu, B.~L., Paz, J.~P., and Zhang, Y., 1992, Phys.\ Rev. {\bf D45}, 2843.

\bibitem{35} Joos, E., 2000, pp. 1-17 in {\it Decoherence: Theoretical, 
Experimental, and  Conceptual Problems}, Ph. Blanchard et al.,  eds. (Springer, Berlin).

\bibitem{36} Joos, E., H. D. Zeh, C. Kiefer, D. Giulini, J. Kupsch, 
I.-O. Stamatescu, 2003, {\it Decoherence and the Appearancs of 
a Classical World in Quantum Theory}, (Springer, Berlin).

\bibitem{37} Kent, A., 1990, Int. J. Mod. Phys. {\bf A5}, 1745.  

\bibitem{38} LaHaye M. D. et al, 2004, Science {\bf 304}, 74.

\bibitem{39} Landauer, R., 1991, Physics Today, {\bf 44}, 23.

\bibitem{40} Laplace, P. S,. 1820, {\it A Philosophical Essay on Probabilities}, English translation of 
the French original by F. W. Truscott and F. L. Emory (Dover, New York 1951).

\bibitem{41} Nielsen, M. A., and I. L. Chuang, 2000, {\it Quantum Computation 
and Quantum Information} (Cambridge University Press).

\bibitem{42} Ollivier, H., et al., 2004, Phys. Rev. Lett. {\bf 93}, 220401.

\bibitem{43} Ollivier, H., et al., 2005, Phys. Rev. {\bf A72}, 423113.

\bibitem{44} Ollivier, H., and W. H. Zurek, 2002, Phys. Rev. Lett. {\bf 88}, 017901.

\bibitem{45} Paz, J.-P., and W. H. Zurek, 2001, p. 533 in {\it Coherent Atomic Matter Waves, 
Les Houches Lectures}, R. Kaiser, C. Westbrook, and F. David, eds.
(Springer, Berlin).

\bibitem{Per} Peres, A., 1993, {\it Quantum Theory: Concepts and Methods} (Kluwer, Dordrecht).

\bibitem{46} Poulin, D., 2005, Phys. Rev. {\bf A71}, 22102.

\bibitem{47} Schr\"odinger, E., 1935, {\it Naturwissenschaften} 807-812; 823-828; 844-849. English translation in Wheeler and Zurek, (1983), p. 152.

\bibitem{48} Squires, E. J., 1990, Phys. Lett. {\bf A145}, 67.

\bibitem{49} Stein, H., 1984, No\^us {\bf 18} 635.

\bibitem{50} Saunders, S., 2004, Proc. Roy. Soc. London, Series A {\bf 460}, 1.

\bibitem{51} Schlosshauer, M. 2004, Rev. Mod. Phys. {\bf 76}, 1267.

\bibitem{52} Schlosshauer, M. 2007, {\it Decoherence} (Springer, Berlin).

\bibitem{53} Schlosshauer, M., and A. Fine,  2005,  {\it Found. Phys.} {\bf 35}, 197.

\bibitem{54} Serafini, A., S. De Siena, F. Illuminati, M.G.A. Paris, 2004, J. Opt. B Quant. Semiclass. Opt. {\bf 6} S591.

\bibitem{55} Tegmark, M., and Shapiro, H.~S., 1994, Phys.\ Rev. {\bf
E50}, 2538.

\bibitem{56} Unruh, W.~G., 1994, Phys. Rev. {\bf A50}, 882.

\bibitem{57} Unruh, W.~G., and Zurek, W.~H., 1989, Phys.\ Rev. {\bf D40}, 1071.

\bibitem{58} von Mises, R., 1939, {\it Probability, Statistics, and Truth} 
(McMillan, New York).

\bibitem{59} von Neumann, J. 1932, {\it Mathematical Foundations of Quantum Theory}, 
translated from German original by R. T. Beyer (Princeton University Press, Princeton, 1955).

\bibitem{60} Wallace, D., 2002, quant-ph/0211104.

\bibitem{61} Wallace, D., 2003, Stud. Hist. Phil. Mod. Phys. {\bf 34}, 415.

\bibitem{Wei} Weissman, M. B., 1999, Found. Phys. Lett. {\bf 12}, 407.

\bibitem{JAW} Wheeler, J. A., 1957, Rev. Mod. Phys. {\bf 29}, 463.

\bibitem{62} Wheeler, J. A., 1983, pp. 182-213 in Wheeler and Zurek, (1983).

\bibitem{63} Wheeler, J. A., 1990, p. 3 in {\it Complexity, Entropy, and the Physics of Information}, Zurek, W. H., ed. (Addison Wesley, Redwood City).

\bibitem{64} Wheeler, J. A., and W. H. Zurek, eds., 1983, {\it Quantum Theory and
Measurement} (Princeton University Press).

\bibitem{65} Wootters, W. K., and W. H. Zurek, 1982, Nature {\bf 299}, 802.

\bibitem{66} Yuen, H. P., 1986, Phys. Lett. {\bf 113A}, 405.

\bibitem{67} Zeh, H. D., 1970, Found. Phys. {\bf 1}, 69, reprinted in Wheeler and Zurek, (1983).

\bibitem{Zeh} Zeh, H. D., 1990, p. 405 in {\it Complexity, Entropy, and the Physics of Information}, Zurek, W. H., ed. (Addison Wesley, Redwood City).

\bibitem{68} Zeh, H. D., 1997, in {\it New Developments on Fundamental Problems 
in Quantum Mechanics}, M. Ferrero and A. van der Merwe, eds. 
(Kluwer, Dordrecht).

\bibitem{Zeh} Zeh, H. D., 2007, {\it The Physical Basis of the Direction of Time} (Springer, Berlin).

\bibitem{69} Zurek, W. H., 1981, Phys. Rev. {\bf D24}, 1516.

\bibitem{70} Zurek, W. H., 1982, Phys. Rev. {\bf D26}, 1862.

\bibitem{71} Zurek, W. H., 1991, Physics Today {\bf 44}, 36; see also
an `update', quant-ph/0306072.

\bibitem{72} Zurek, W. H., 1993, Progr. Theor. Phys. {\bf 89}, 281.

\bibitem{73} Zurek, W. H.,  1998a, Physica Scripta {\bf T76}, 186.

\bibitem{74} Zurek, W. H.,  1998b, Phil. Trans. Roy. Soc. London, Series A 
{\bf 356}, 1793.

\bibitem{75} Zurek, W. H., 2003a, Rev. Mod. Phys. {\bf 75}, 715.

\bibitem{76} Zurek, W. H., 2003b, Phys. Rev. Lett. {\bf 90}, 120404.

\bibitem{77} Zurek, W. H., 2003c, Phys. Rev. {\bf A67}, 012320.

\bibitem{78} Zurek, W. H., 2005, Phys. Rev. {\bf A71}, 052105.

\bibitem{79} Zurek, W. H., 2007, arXiv:quant-ph/0703160.

\bibitem{80} Zurek, W.~H., Habib, S., and Paz, J.-P., 1993, Phys.\
Rev.\ Lett. {\bf 70}, 1187.

\end{thebibliography}
\end{document}